\documentclass[twocolumn]{aastex61}
\usepackage{natbib}
\hypersetup{linkcolor=red,citecolor=blue,filecolor=cyan,urlcolor=magenta}
\newcommand{\kms}{km~s$^{-1}$}
\newcommand{\mstar}{\ensuremath{\mathcal{M}_{*}}}
\newcommand{\mhalo}{\ensuremath{\mathcal{M}_{h}}}
\newcommand{\msun}{\ensuremath{\mathcal{M}_{\odot}}}
\newcommand{\sigmasfr}{\Sigma_{\textnormal{\scriptsize{SFR}}}}
\newcommand{\units}{M$_{\odot}$~yr$^{-1}$~kpc$^{-2}$}
\newcommand{\sigunits}{M$_{\odot}$~kpc$^{-2}$}
\newcommand{\myr}{M$_{\odot}$~yr$^{-1}$}
\newcommand{\nev}{[\textrm{Ne}~\textsc{v}]}
\newcommand{\mgii}{\textrm{Mg}~\textsc{ii}}

\newcommand{\oiii}{[\textrm{O}~\textsc{iii}]}
\newcommand{\oii}{[\textrm{O}~\textsc{ii}]}

\newcommand{\lrest}{\lambda_{\textnormal{\scriptsize{rest}}}}
\newcommand{\etal}{et~al.}
\newcommand{\prospector}{\texttt{Prospector}}
\newcommand{\astropy}{\texttt{Astropy}}
\newcommand{\dynesty}{\texttt{dynesty}}
\newcommand{\galfit}{\texttt{GALFIT}}
\newcommand{\galfitm}{\texttt{GALFITM}}
\newcommand{\fsps}{\texttt{FSPS}}

\accepted{for publication in ApJ}
\shorttitle{Compact Starburst Galaxies with Fast Outflows}
\shortauthors{Diamond-Stanic \etal}
\begin{document}

\title{Compact Starburst Galaxies with Fast Outflows: Central Escape Velocities \\ and Stellar Mass Surface Densities from Multi-band Hubble Space Telescope Imaging}

\correspondingauthor{Aleksandar M. Diamond-Stanic}
\email{adiamond@bates.edu}

\author{Aleksandar M. Diamond-Stanic}
\affil{Department of Physics and Astronomy, Bates College, Lewiston, ME, 04240, USA}

\author{John Moustakas}
\affil{Department of Physics and Astronomy, Siena College, Loudonville, NY 12211, USA}

\author{Paul H. Sell}
\affil{Department of Astronomy, University of Florida, Gainesville, FL, 32611 USA}

\author{Christy A. Tremonti}
\affil{Department of Astronomy, University of Wisconsin-Madison, Madison, WI 53706, USA}

\author{Alison L. Coil}
\affil{Center for Astrophysics and Space Sciences, University of California, San Diego, La Jolla, CA 92093, USA}

\author{Julie D. Davis}
\affil{Department of Astronomy, University of Wisconsin-Madison, Madison, WI 53706, USA}

\author{James E. Geach}
\affil{Centre for Astrophysics Research, University of Hertfordshire, Hatfield, Hertfordshire AL10 9AB, UK}

\author{Sophia C.~W. Gottlieb}
\affil{Department of Physics and Astronomy, Bates College, Lewiston, ME, 04240, USA}

\author{Ryan C. Hickox}
\affil{Department of Physics and Astronomy, Dartmouth College, Hanover, NH 03755, USA}

\author{Amanda Kepley}
\affil{National Radio Astronomy Observatory, Charlottesville, VA 22903, USA}

\author{Charles Lipscomb}
\affil{Department of Physics and Astronomy, Bates College, Lewiston, ME, 04240, USA}
\affil{Department of Aerospace Engineering, University of Colorado, Boulder, CO, 80303, USA}

\author{Joshua Rines}
\affil{Department of Physics and Astronomy, Bates College, Lewiston, ME, 04240, USA}
\affil{Department of Earth and Environmental Sciences, Boston College, Chestnut Hill, MA, 02467, USA}

\author{Gregory H. Rudnick}
\affil{Department of Physics and Astronomy, University of Kansas, Lawrence, KS 66045, USA}

\author{Cristopher Thompson}
\affil{Department of Physics and Astronomy, Bates College, Lewiston, ME, 04240, USA}
\affil{Department of Physics, Ohio State University, Columbus, OH, 43210, USA}

\author{Kingdell Valdez}
\affil{Department of Physics and Astronomy, Bates College, Lewiston, ME, 04240, USA}

\author{Christian Bradna}
\affil{Department of Physics and Astronomy, Bates College, Lewiston, ME, 04240, USA}

\author{Jordan Camarillo}
\affil{Department of Physics and Astronomy, Bates College, Lewiston, ME, 04240, USA}
\affil{Department of Biomedical Engineering, Boston University, Boston, MA, 02215, USA}

\author{Eve Cinquino}
\affil{Department of Physics and Astronomy, Bates College, Lewiston, ME, 04240, USA}
\affil{Physical Oceanography Department, Woods Hole Oceanographic Institute, Woods Hole, MA, 02543, USA}

\author{Senyo Ohene}
\affil{Department of Physics and Astronomy, Bates College, Lewiston, ME, 04240, USA}

\author{Serena Perrotta}
\affil{Center for Astrophysics and Space Sciences, University of California, San Diego, La Jolla, CA 92093, USA}

\author{Grayson C. Petter}
\affil{Department of Physics and Astronomy, Dartmouth College, Hanover, NH 03755, USA}

\author{David S.~N. Rupke}
\affil{Department of Physics, Rhodes College, Memphis, TN, 38112, USA}

\author{Chidubem Umeh}
\affil{Department of Physics and Astronomy, Bates College, Lewiston, ME, 04240, USA}

\author{Kelly E. Whalen}
\affil{Department of Physics and Astronomy, Dartmouth College, Hanover, NH 03755, USA}

\begin{abstract}
We present multi-band Hubble Space Telescope imaging that spans rest-frame near-ultraviolet through near-infrared wavelengths ($\lrest=0.3$--1.1~$\mu$m) for 12 compact starburst galaxies at $z=0.4$--0.8. These massive galaxies ($\mstar\sim10^{11}~\msun$) are driving very fast outflows ($v_{max}=1000$--3000~\kms), and their light profiles are dominated by an extremely compact starburst component ($\textnormal{half-light radius}\sim100$~pc). Our goal is to constrain the physical mechanisms responsible for launching these fast outflows by measuring the physical conditions within the central kiloparsec. Based on our stellar population analysis, the central component typically contributes $\approx$25\% of the total stellar mass and the central escape velocities $v_{esc,central}\approx900$~\kms\ are a factor of two smaller than the observed outflow velocities. This requires physical mechanisms that can accelerate gas to speeds significantly beyond the central escape velocities, and it makes clear that these fast outflows are capable of traveling into the circumgalactic medium, and potentially beyond. We find central stellar densities $\Sigma_{e,central}\approx3\times10^{11}$~\sigunits\ comparable to theoretical estimates of the Eddington limit, and we estimate $\Sigma_1$ surface densities within the central kpc comparable to those of compact massive galaxies at $0.5<z<3.0$. Relative to ``red nuggets" and ``blue nuggets" at $z\sim2$, we find significantly smaller $r_e$ values at a given stellar mass, which we attribute to the dominance of a young stellar component in our sample and the better physical resolution for rest-frame optical observations at $z\sim0.6$ versus $z\sim2$. We compare to theoretical scenarios involving major mergers and violent disc instability, and we speculate that our galaxies are progenitors of power-law ellipticals in the local universe with prominent stellar cusps.  
\end{abstract}

%\keywords{galaxies: evolution --- galaxies: starburst}

\section{Introduction} \label{sec:intro}

The cosmic inefficiency of star formation across a wide range of mass scales for galaxies and dark matter halos is implied by measurements of the baryon density \citep[e.g.,][]{hin13,pla15,coo18} and the galaxy stellar mass function \citep[e.g.,][]{bel03,mos10,mou13}. The fact that only $\sim5\%$ of baryonic matter has formed stars by the present is often explained by feedback process that inject energy and momentum into interstellar and circumgalactic gas that would otherwise cool and collapse to form stars \citep[e.g.,][]{som15}. Consistent with this hypothesis, observations of large-scale outflows from star-forming galaxies and active galactic nuclei (AGNs) provide empirical evidence for ejective feedback as a mechanism for accelerating, heating, and potentially removing the interstellar gas supply \citep[e.g.,][]{vei05,fab12,vei20}.

Regarding the physical mechanisms responsible for driving large-scale outflows, the observed scaling relationships between outflow properties and galaxy properties provide some insight. For example, it is well known that outflow velocity scales with both stellar mass and star-formation rate (SFR) when considering samples of galaxies that span several orders of magnitude in terms of dynamic range \citep[e.g.,][]{mar05,rup05,chi15,hec15}. This has provided justification for feedback models in which the outflow velocity is tied to the galaxy rotation speed, velocity dispersion, escape velocity, or dark matter virial velocity \citep[e.g.,][]{mur05,opp06,mur11,tho15}. It is worth noting that trends between outflow velocity and galaxy properties have not been readily apparent when galaxies from an individual survey are considered on their own \citep[e.g.,][]{kor12,rub14}, which illustrates the significant scatter in these correlations. That said, when including galaxies from the literature that span more extreme limits for galaxy physical parameters, there is also evidence that outflow velocity scales with the surface density of star formation \citep[e.g.,][]{law12,kor12,hec16,pet20} in the sense that faster outflows tend to be associated with the galaxies that have more intense and compact star formation.  

Regarding the formation and assembly of massive galaxies, it has been known for more than a decade based on observations with the Hubble Space Telescope (HST) that compact morphologies are quite common among massive quiescent galaxies at $z\sim2$ \citep[e.g.,][]{dad05,tru06,zir07,van08,bui08}. Furthermore, studies of compact, star-forming galaxies at $z\sim3$ \citep[e.g.,][]{bar13,pat13,ste13,wil14}, which are likely progenitors of compact quiescent galaxies at $z\sim2$, have found evidence for outflows and ongoing AGN activity \citep[e.g.,][]{ran14,wil15,koc17}. This suggests an evolutionary connection between events that dissipate the angular momentum of the cold gas supply in massive galaxies, triggering intense and compact star formation \citep[e.g.,][]{mih96,dek14}, and the subsequent quenching of star formation, perhaps driven by large-scale outflows. This motivates further study of the assembly of such galaxies and the physical processes responsible for depleting and ejecting their cold gas supply. 

We have been studying a sample of compact massive galaxies with clear evidence for recent gas depletion and ejection that may provide insight more broadly into the formation of compact quiescent galaxies and the physical mechanisms responsible for driving powerful outflows that are capable of quenching star formation \citep{tre07,dia12,gea13,gea14,sel14,gea18,rup19,pet20}. Following the spectroscopic discovery of high-velocity outflows in this sample of galaxies at $z=0.4$--0.8, which also exhibit spectral characteristics akin to post-starburst galaxies \citep{tre07}, we discovered that these galaxies have remarkably compact optical morphologies with signatures of major mergers \citep{dia12,sel14} and infrared (IR) and ultraviolet (UV) luminosities that indicate incredibly high SFR surface densities \citep{dia12}. Furthermore, our observations of CO transitions at millimeter wavelengths have provided evidence for short gas depletion timescales and fast, large-scale outflows of molecular gas \citep{gea13,gea14,gea18,rup19}. Most galaxies in the sample show no evidence for AGN activity based on X-ray observations, optical emission lines, and infrared colors, and for the galaxies with evidence for ongoing black hole accretion, the AGN emission contributes a small fraction of the bolometric luminosity \citep{dia12,sel14}. Taken as a whole, these multi-wavelength observations paint a picture of massive galaxies that formed via highly dissipative, gas-rich mergers and are exhausting their gas supply through a violent process of compact star formation and extreme feedback.  

In this context, an important and outstanding question about the galaxies in this sample has been the nature of their extreme compactness. In particular, the question of how compact the stellar mass, as opposed to the light, is for these galaxies has important implications for testing feedback models, some of which predict that the outflow velocity should be comparable to the escape velocity \citep[e.g.,][]{mur05,opp06,mur11} and some of which are capable of producing outflow velocities that exceed the central escape velocity by a significant factor \citep[e.g.,][]{spr03,hec11,tho15,bus16}. Is the compactness of their light profiles a result of their optical light being dominated by a recent starburst \citep[which might contribute $10\%$--30\% to the total stellar mass,][]{hop09} or are their stellar mass profiles similarly compact? The answer to this question requires an empirical measurement of spatially resolved colors on sub-kpc scales and robust stellar population modeling for our compact galaxy sample. In this paper, we focus on measurements of the spatial profile in multiple bands taken with HST imaging and the interpretation of those results. 

We describe this sample of galaxies in more detail in Section~\ref{sec:sample} and we describe our HST imaging data in Section~\ref{sec:data}. In Section~\ref{sec:photmorph}, we present an analysis of the morphology and multi-band photometry of the compact starburst components of these galaxies, and in Section~\ref{sec:stellarpop} we present stellar population modeling of the central, extended, and total spectral energy distributions (SEDs). We discuss the implications of these results for central escape velocity, stellar mass surface density, and theoretical models in Section~\ref{sec:escape_surface}, and we conclude in Section~\ref{sec:summary}. For calculations of luminosity, stellar mass, and angular size, we adopt a cosmology with $\Omega_{\Lambda}=0.7$, $\Omega_{M}=0.3$, and $h=H_0/(\textnormal{100~km~s}^{-1}~\textnormal{Mpc}^{-1})=0.7$. For stellar mass calculations, we use a \citet{sal55} initial mass function. 

\begin{deluxetable*}{lcrcccc}
\tablecaption{Sample properties \label{tab:sample}}
\tablehead{
\colhead{Galaxy} & \colhead{RA} & \colhead{Dec} & \colhead{$z$} & \colhead{$v_{max}$} & \colhead{$v_{avg}$} & \colhead{$\log(\mstar)$} \\
\colhead{(SDSS)} & \colhead{[degrees]} & \colhead{[degrees]} & \colhead{} & \colhead{[\kms]} & \colhead{[\kms]} & \colhead{$[\msun]$}  
}
\decimalcolnumbers
\startdata
J0826+4305\tablenotemark{a} & 126.66006 & 43.09150 & 0.603 & $-1660$ & $-1250$ & 10.90 \\ %11.03 \\%& 2.20 \\
J0901+0314 & 135.38926 & 3.23680 & 0.459 & $-1780$ & $-1310$ & 10.81 \\ %10.98 \\%& 1.93 \\
J0905+5759 & 136.34832 & 57.98679 & 0.712 & $-3020$ & $-2510$ & 10.98 \\ %10.74 \\%& 2.18 \\
J0944+0930\tablenotemark{a} & 146.07437 & 9.50539 & 0.514 & $-1990$ & $-1290$ & 10.80 \\ %10.69 \\%& 1.92 \\
J1107+0417 & 166.76197 & 4.28410 & 0.467 & $-2050$ & $-1420$ & 10.89 \\ %10.73 \\%& 1.65 \\
J1219+0336 & 184.98241 & 3.60442 & 0.451 & $-2030$ & $-1680$ & 11.11 \\ %10.71 \\%& 2.16 \\
J1341$-$0321 & 205.40333 & $-3.35702$ & 0.658 & $-2000$ & $-760$ & 10.86 \\ %10.98 \\%& 2.07 \\
J1506+5402\tablenotemark{b,c} & 226.65124 & 54.03909 & 0.608 & $-2450$ & $-1290$ & 10.84 \\ %10.96 \\%& 2.22 \\
J1558+3957\tablenotemark{a} & 239.54683 & 39.95579 & 0.402 & $-1260$ & $-870$ & 10.77 \\ %10.91 \\%& 1.94 \\
J1613+2834\tablenotemark{b} & 243.38552 & 28.57077 & 0.449 & $-2500$ & $-810$ & 11.13 \\ %10.98 \\%& 2.13 \\
J2116$-$0634 & 319.10479 & $-6.57911$ & 0.728 & $-2240$ & $-1120$ & 11.11 \\ %10.92 \\%& 2.24\\
J2140+1209\tablenotemark{a,d} & 325.00206 & 12.15405 & 0.752 & $-1040$ & $-510$ & 11.16 \\ %11.28 \\%& 2.37 \\
\enddata
\tablecomments{Column 1: Abbreviated SDSS IAU designation. Column 2: Right ascension in degrees. Column 3: Declination in degrees. Column 4: redshift. Column 5: Maximum outflow velocity, defined by the point at which the equivalent width distribution for $\mgii~\lambda2796$ reaches 95\% of the total. Column 6: Average outflow velocity, defined by the median of the cumulative equivalent width distribution for $\mgii~\lambda\lambda2796,2803$. Column 7: Total stellar mass, measured as described in Section~\ref{sec:extended}.}
\tablenotetext{a}{This galaxy has an X-ray upper limit from pointed X-ray observations by \citet{sel14} (see text for details).}
\tablenotetext{b}{This galaxy has a weak X-ray detection from \citet{sel14} that is consistent with its star-formation rate (see text for details).}
\tablenotetext{c}{This galaxy has emission-line detections and ratios consistent with either a starburst or a sub-dominant obscured AGN (see text for details).}
\tablenotetext{d}{This galaxy has a weak broad \mgii\ line, indicating a type 1 AGN component that is sub-dominant in the HST bands (see text for details).}
\end{deluxetable*}

\begin{figure*}[!t]
\begin{center}
\includegraphics[angle=0,scale=0.85]{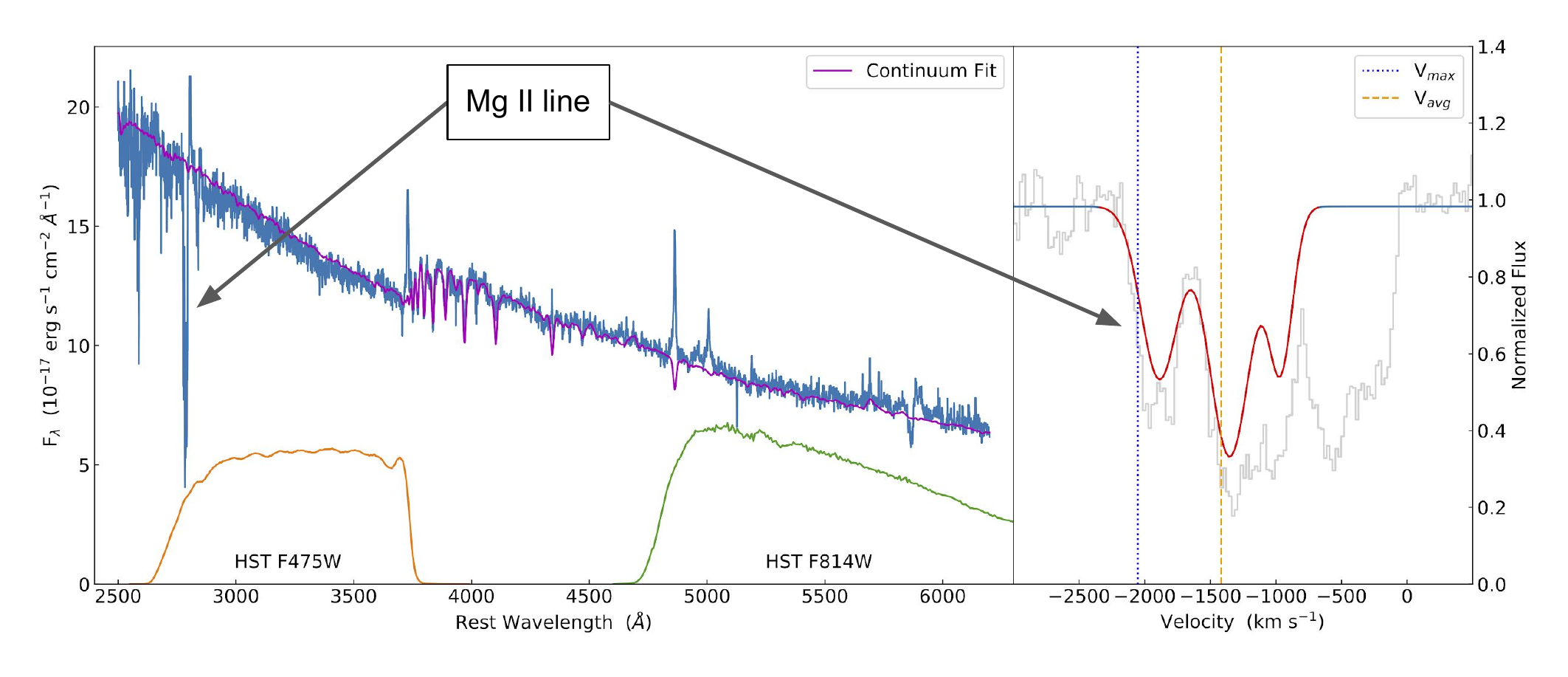}
\caption{Left: The rest-frame near-UV through optical spectrum ($\lrest=2500$--6200~\AA) for the galaxy J1107+0417 with a continuum fit overlaid. The throughput curves for 2/3 of the HST filters used in this paper are shown beneath the spectrum; the F475W filter traces near-UV emission below the Balmer break and the F814W filter traces optical emission. The prominent absorption line near $\lrest=2800$~\AA\ is interstellar \mgii. Right: The $\mgii~\lambda\lambda2796,2803$ absorption-line profile in velocity space. The zero of the velocity axis corresponds to the expected location of the $\mgii~\lambda2796$ line based on the systemic redshift of the galaxy determined from the stellar absorption lines in the continuum fit. The two lines of the \mgii\ doublet overlap for this galaxy, and a model fit including three velocity components is shown to represent the intrinsic gas velocity profile (Davis \etal, in prep). The $v_{avg}=-1420$~\kms\ value (marked by the yellow dashed line) corresponds to the median of the equivalent width distribution for $\mgii~\lambda2796$, and the $v_{max}=-2050$~\kms\ value (marked by the blue dotted line) is defined by the point at which the equivalent width distribution reaches 95\% of the total.}
\label{fig:spectrum}
\end{center}
\end{figure*}

\section{Sample Selection}\label{sec:sample}

The parent sample of galaxies for this study was selected based on spectroscopy from the Sloan Digital Sky Survey \citep[SDSS,][]{yor00} as described previously by \citet{tre07}, \citet{dia12}, and \citet{sel14}. In particular, this parent sample includes $1089$ galaxies at $0.35<z<1$ from the Seventh Data Release \citep[SDSS DR7,][]{aba09} that have young stellar populations dominated by B- and A-stars and moderately weak nebular emission (\oii\ rest-frame equivalent width $<20$~\AA). From this parent sample, we identified a sub-sample of 131/1089 galaxies (the ``follow-up sample") using additional cuts based on redshift ($z>0.4$ so the \mgii~$\lambda\lambda2796,2803$ doublet would be easily accessible with optical spectrographs), emission-line strength (\oii\ rest-frame equivalent width $<15$~\AA), and optical magnitude ($g<20$ and $i<20$). These galaxies have been the focus of subsequent follow-up observations, including ground-based spectroscopy with the MMT, Magellan, and Keck \citep[$N=50$ galaxies observed to date;][]{tre07,dia12,sel14,gea14,dia16,gea18,rup19}, X-ray imaging with Chandra \citep[$N=12$ galaxies;][]{sel14}, and optical imaging with the Hubble Space Telescope \citep[$N=29$ galaxies;][]{dia12,sel14,gea14,dia16,gea18,rup19}.

The 29/131 galaxies from the follow-up sample that also have HST observations (the ``HST sample") were targeted in Cycle 17 ($N=12$ galaxies, Program ID \href{http://www.stsci.edu/cgi-bin/get-proposal-info?id=12019}{12019}, PI: Tremonti) or Cycle 18 ($N=17$ galaxies, Program ID \href{http://www.stsci.edu/cgi-bin/get-proposal-info?id=12272}{12272}, PI: Tremonti) with the Wide Field Camera 3 \citep[WFC3,][]{kim08} using the UVIS channel and the F814W filter. The Cycle 17 program included joint X-ray observations with Chandra, and it focused on the 12 galaxies with the strongest evidence for ongoing AGN activity based on emission-line properties. These HST and Chandra observations were published by \citet{sel14}. The Cycle 18 program focused on the galaxies that had the youngest derived ages for the recent burst of star formation ($t_{\textnormal{\scriptsize{burst}}}<300$~Myr) based on follow-up spectroscopy. This selection yielded a sample of galaxies with bluer colors and stronger emission lines than typically found in post-starburst galaxy samples, and our subsequent analysis has revealed evidence for significant ongoing star formation ($>50$~\myr) in 14/29 of these galaxies based on WISE mid-infrared luminosities \citep{dia12}. 

The 12/29 galaxies from the HST sample that we focus on for this paper (see Table~\ref{tab:sample}) include the galaxies with the largest SFR surface densities measured by \citet{dia12}, spanning the range $30$~\units$<\sigmasfr<3000$~\units. This range in SFR surface density reflects some variation across the sample, with the top half (6/12) all having $\textnormal{SFR}>100$~\myr, $r_e<200$~pc, and $\sigmasfr>300$~\units, while the bottom half includes galaxies that are somewhat less extreme (e.g., J1558$+$3957, an ongoing merger for which the total light is not quite so dominated by a single source, and J1613$+$2834, for which the central component is significantly less compact with $r_e\approx0.7$~kpc). Half of these galaxies (6/12) were included in the sample published by \citet{sel14}, one quarter (3/12) have published molecular gas observations \citep{gea13,gea14,gea18}, and three quarters (9/12) have published radio measurements at 1.5~GHz \citep{pet20}. 

None of the galaxies in this paper would be classified as an AGN on the basis of WISE mid-infrared colors; they do not meet the $\textrm{W1}-\textrm{W2}>0.8$ (Vega) threshold from \citet{ste12} or the reliable AGN criteria from \citet{ass13}. There are two galaxies with weak X-ray detections (J1506+5402 and J1613+2834, 4 X-ray counts each) from \citet{sel14} that indicate $L_X\approx10^{42}$~erg~s$^{-1}$ and four galaxies with X-ray upper limits (J0826+4305, J0944+0930, J1558+3957, J2140+1209) that imply $L_X<10^{43}$~erg~s$^{-1}$. All of these galaxies are consistent with the relationship between X-ray luminosity and mid-infrared luminosity for starburst galaxies \citep{asm11,min14,sel14}. One of the galaxies with a weak X-ray detection (J1506+5402) also has a clear detection of \nev~$\lambda3426$ and an emission-line ratio $\oiii/\textrm{H}\beta\approx1$, which is the highest ratio in the sample and consistent with a composite BPT classification \citep{bal81,kew01,kau03,sel14}. These \nev\ and \oiii\ emission lines could be produced either by an extreme ($t\sim3$~Myr) starburst or an AGN that contributes $\approx10$\% of the mid-infrared continuum \citep{dia12,gea13,sel14}. The galaxy with the slowest outflow velocity in the sample (J2140+1209) has a weak broad \mgii\ line, which is a clear indication of a type 1 AGN component, and spectral modeling suggests a $\approx20$\% AGN continuum contribution at $\lrest=2800$~\AA\ \citep{sel14}; the other 11/12 galaxies show no evidence for any AGN continuum contributions at near-ultraviolet or optical wavelengths on the basis of high signal-to-noise spectroscopy (as described above and shown in Figure~\ref{fig:spectrum}, which we describe below). In summary, there is some evidence for AGN activity in several of these 12 galaxies, but in all cases the upper limits on the AGN emission indicate that it contributes a small fraction of the galaxy's bolometric luminosity.  

All 12 galaxies were targeted in Cycle 22 (Program ID \href{http://www.stsci.edu/cgi-bin/get-proposal-info?id=13689}{13689}, PI: Diamond-Stanic) with WFC3 in two additional bands, using the UVIS channel with the F475W filter and the IR channel with the F160W filter. When combined with the existing WFC3 UVIS/F814W images, which probe the rest-frame optical, these 12 galaxies have high-resolution imaging at rest-frame near-ultraviolet through rest-frame near-infrared wavelengths. While spectroscopy is not the focus of this paper, all 12 galaxies also have very fast outflows with $|v_{max}|=1000$--3000~\kms\ as measured from absorption-line spectroscopy of the \mgii~$\lambda\lambda2796,2803$ doublet (Davis \etal, in prep). In Figure~\ref{fig:spectrum}, we show an example spectrum for the galaxy J1107+0417 over the wavelength range $\lrest=2500$--6200~\AA\ based on data from the Magellan/MagE spectrograph \citep{mar08}. We highlight the spectral regions on opposite sides of the Balmer break that correspond to the F475W and F814W filters, and we show the gas velocity profile based on the \mgii~$\lambda2796$ absorption line. The outflow velocities in Table~\ref{tab:sample} are measured from this line, with $v_{avg}$ based on the median of the equivalent width distribution and $v_{max}$ based on the point at which the equivalent width distribution reaches 95\% of the total. Similar spectra for the galaxies in this paper have been published by \citet{tre07} (see their Figure~1, which includes two galaxies from this paper), \citet{dia12} (see their Figure~3, which includes three galaxies from this paper), and \citet{sel14} (see their Figure~2, which includes six galaxies from this paper).

\section{Data}\label{sec:data}

As described in Section~\ref{sec:sample}, the 12 galaxies in this paper have WFC3 UVIS/F814W observations from Cycle 17 or Cycle 18 and more recent Cycle 22 observations with WFC3 UVIS/F475W and WFC3 IR/F160W. For each filter, the data were obtained using four exposures in a single orbit, with exposure times ranging from 10-12 minutes per exposure (i.e., at least 40 minutes of exposure time per filter per galaxy). We used sub-pixel dither patterns to reject hot pixels and cosmic rays and to improve spatial resolution. At the mean redshift of our sample ($z=0.57$), the central wavelengths of these filters correspond to $\lrest\textnormal{(F475W)}\approx3000$~\AA, $\lrest\textnormal{(F814W)}\approx5200$~\AA, and $\lrest\textnormal{(F160W)}\approx1.02~\mu$m. Our immediate observational goal with these multi-band observations is to measure rest-frame ultraviolet, optical, and near-infrared colors in a spatially resolved manner on the smallest possible angular scales. 

In order to achieve this goal, we re-processed the F814W data that were originally published by \citet{dia12} and \citet{sel14} along with our new analysis of the F475W and F160W images. By doing so, we are able to produce images that are matched on a pixel-by-pixel level in the final science frames used for photometric measurements. Specifically, we begin with the four calibrated exposures for each filter and use the DrizzlePac software to align and combine images for each filter. For the UVIS data, we use the calibrated exposures that have been processed by the CALWF3 pipeline, which includes a pixel-based correction for Charge Transfer Efficiency (CTE). As a first step, we use the astrodrizzle function to produce an initial stacked image for each filter. As a second step, we use the tweakreg function to align the F475W and F160W images to the F814W image. After using the tweakback function to update the header astrometry for the F475W and F160W calibrated individual exposures, we then run astrodrizzle a final time to produce final images in F475W and F160W that are aligned on a pixel-by-pixel level with the F814W images. 

An important point is that the pixel scale and PSF FWHM are significantly different for the UVIS and IR channels. The UVIS images have a native pixel size of 0.04$\arcsec$/pixel and the IR images have a native pixel size of 0.13$\arcsec$/pixel. Furthermore, the PSF FWHM values range from 0.07$\arcsec$ for UVIS (F475W, F814W) to 0.15$\arcsec$ for IR (F160W). The UVIS/F475W PSF (0.067$\arcsec$) is also slightly narrower than the UVIS/F814W PSF (0.074$\arcsec$).  We use a plate scale of 0.05$\arcsec$/pixel when creating pixel-aligned images in all three filters, which is a compromise between sampling the UVIS PSF with a pixel scale that is smaller than the UVIS PSF FWHM while not oversampling the IR PSF. When considering the UVIS data on their own, we run the same DrizzlePac functions to create separate stacked images with a plate scale of 0.025$\arcsec$/pixel. 

To facilitate photometric analysis and model fitting, we also generate uncertainty images for each science image. We begin with the stacked images produced by DrizzlePac, and we estimate the noise in each pixel by accounting for shot noise in the background-subtracted flux, shot noise in the background, read noise, and dark current noise. We estimate shot noise in the background-subtracted flux for each pixel by taking the square root of the counts in units of electrons. We estimate shot noise in the background by fitting a Gaussian to a histogram of pixel values across the entire image, excluding those pixels at the upper end of the histogram that are dominated by source flux. We use values for read noise and dark current noise from the image headers and the WFC3 Instrument Handbook. This method for estimating the uncertainty in each pixel does not account for additional noise co-variance between pixels introduced by the drizzling process, but it does provide a useful baseline for evaluating two-dimensional model fits in our subsequent analysis.

\begin{figure*}[!t]
\begin{center}
\includegraphics[angle=0,scale=0.5]{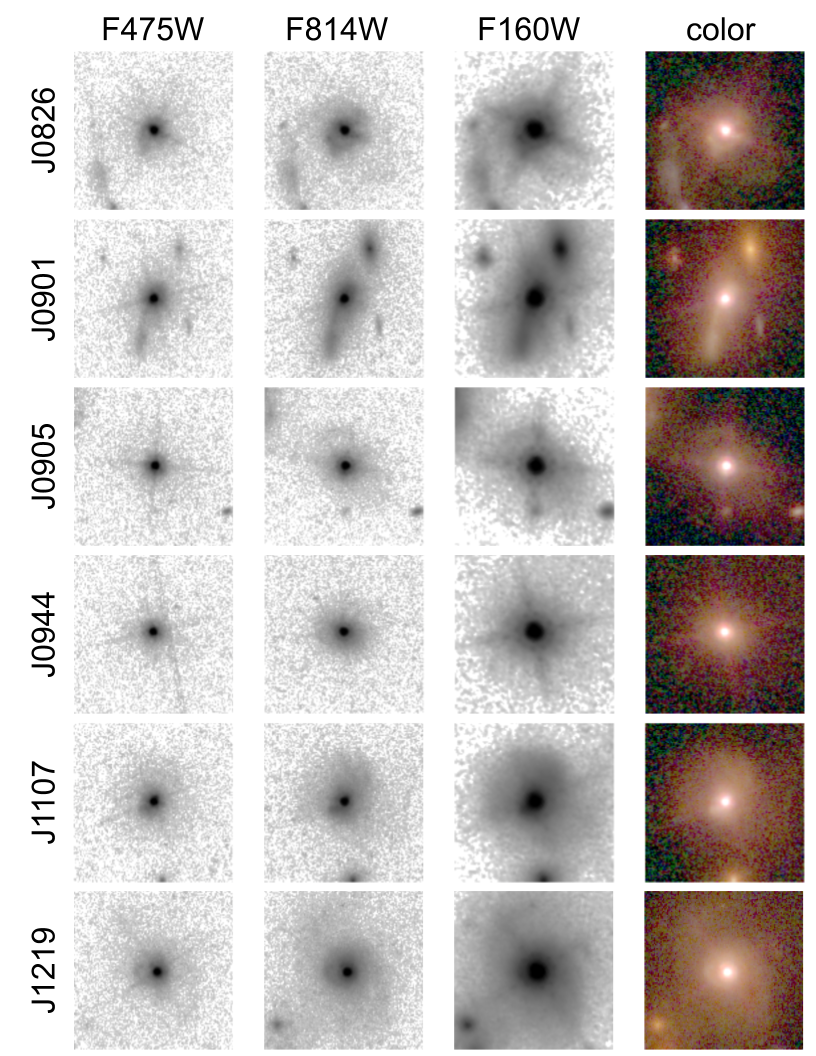}
\caption{Left to right: HST/WFC3 images in the UVIS/F475W, UVIS/F814W, and IR/F160W bands for 6/12 galaxies. The three images for each galaxy are combined in the rightmost column to produce a blue-green-red color composite. Each image is 6$\arcsec$ across, which corresponds to $\approx40$~kpc at the median redshift of the sample, and is oriented such that north is up and east is to the left. Top to bottom: images for J0826, J0901, J0905, J0944, J1107, and J1219. }
\label{fig:hst_images_1}
\end{center}
\end{figure*}

\begin{figure*}[!th]
\begin{center}
\includegraphics[angle=0,scale=0.5]{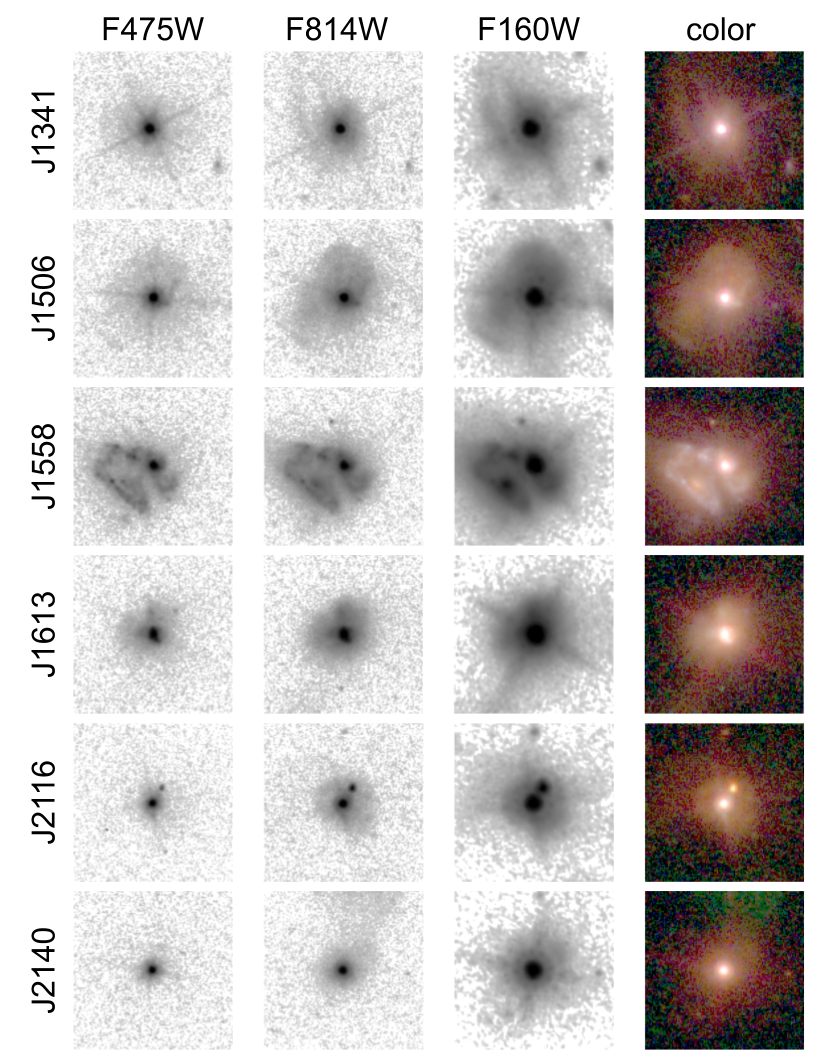}
\caption{Left to right: HST/WFC3 images in the UVIS/F475W, UVIS/F814W, and IR/F160W bands for 6/12 galaxies. The three images for each galaxy are combined in the rightmost column to produce a blue-green-red color composite. Each image is 6$\arcsec$ across, which corresponds to $\approx40$~kpc at the median redshift of the sample, and is oriented such that north is up and east is to the left. Top to bottom: images for J1341, J1506, J1558, J1613, J2116, and J2140.}
\label{fig:hst_images_2}
\end{center}
\end{figure*}

\section{Morphological and Photometric Analysis}\label{sec:photmorph}

As described by \citet{dia12} and \citet{sel14}, a key challenge in quantifying the photometric and morphological characteristics of these galaxies based on HST imaging is that the half-light radii inferred from single-component models with \galfit\ \citep{pen02,pen10} are often comparable to, or even smaller than, the width of a single pixel. In other words, the two-dimensional light profiles of these galaxies on scales smaller than a kiloparsec are dominated by the width and shape of the point-spread function. This can be seen in Figure~\ref{fig:hst_images_1} and Figure~\ref{fig:hst_images_2} in which the central source dominates the central pixels in each image. These images are $6\arcsec$ on a side, whereas the median half-light radius measured from the UVIS images is $0.02\arcsec$ (see Section~\ref{sec:uvis_re} and Table~\ref{tab:re}). For comparison, the pixel values in the central $0.2\arcsec\times0.2\arcsec$ region of each F475W and F814W image ($\textnormal{FWHM}\approx0.07\arcsec$) and the central $0.4\arcsec\times0.4\arcsec$ of each F160W image ($\textnormal{FWHM}\approx0.15\arcsec$) exceed the maximum scale parameter used to produce the figures (i.e., these pixels are completely black in Figure~\ref{fig:hst_images_1} and Figure~\ref{fig:hst_images_2}). Therefore, our desire to understand the intrinsic light profile down to $\sim0.1$~kpc scales requires the use of an empirically-generated PSF and modeling of the observed light profile to infer information about regions that are smaller than the PSF FWHM. 

Regarding photometric measurements, our goal is to perform photometry on the same morphological component in each image: the compact central starbursts that were found by \citet{dia12} and \citet{sel14} to dominate the light profile in rest-frame optical images and inferred to have sizes as small as $r_e\sim100$~pc. From that perspective, our morphological and photometric analyses of the central source in each galaxy are driven by the goals of (1) finding an accurate size estimate for the central source, and (2) measuring the flux of this source on the same physical scale in all three images. 

Regarding the estimate of the half-light radius, we focus on the UVIS bands (F475W, F875W), which have the best spatial resolution and which are the most clearly dominated by the central source (see Section~\ref{sec:uvis_re}). The increased prominence of non-nuclear, extended emission at longer wavelengths is clear from Figure~\ref{fig:hst_images_1} and Figure~\ref{fig:hst_images_2}. This is a signature of an extended stellar population that is older and redder than the central starburst (see Section~\ref{sec:extended}). To facilitate measurements on the same physical scale in each image, we perform a simultaneous fit to all three photometric bands, using the half-light radius for the central component measured from the UVIS bands (see Section~\ref{sec:uvis_re}) as a fixed constraint for all images (see Section~\ref{sec:galfitm}).

\subsection{Measuring the half-light radius of the central component with \galfit}\label{sec:uvis_re}

For our measurements of half-light radius, we perform fits to both the F475W and F814W images using the \galfitm\ software developed by the MegaMorph collaboration \citep{hau13,vik13}. This software creates model images in multiple bands and then convolves each model image with a PSF before comparing with the data, minimizing residuals using the Levenberg-Marquardt method \citep{pen02}. To generate the empirical PSF for each image, we use stars from the same image and the method described by \citet{sel14}. For this analysis, we use the $0.025\arcsec$/pixel images, which have the best spatial sampling of the central source. Following \citet{dia12} and \citet{sel14}, we use single-component S{\'e}rsic profiles with $n=4$ \citep{dev48,ser63}. To quantify the values and uncertainties for half-light radii ($r_e$) and magnitudes ($m_{\textnormal{\scriptsize{F475W}}}$, $m_{\textnormal{\scriptsize{F814W}}}$), we generate a suite of models with \galfitm\ over a grid of $r_e$, $m_{\textnormal{\scriptsize{F475W}}}$, and $m_{\textnormal{\scriptsize{F814W}}}$ for each galaxy. To provide initial estimates for these values, we perform simultaneous fits to the F475W and F814W images using a square image region that is $200\times200$ pixels or $5\arcsec$ on a side; this corresponds to $\sim30$~kpc at the median redshift of our sample, so this initial region encompasses the vast majority of the emission for each galaxy. For this multi-band analysis, we use the same $r_e$, position angle, and axis ratio values for both the F475W and F814W images.

For our subsequent analysis, we focus on the central region ($0.5\arcsec\times0.5\arcsec$ or $\approx3~\textnormal{kpc}\times3$~kpc) of each galaxy for which the diffraction pattern from the central source dominates the light profile. This allows the data vs model comparison to be based on pixels that do not have significant contributions from tidal and other extended features. For our model comparisons, we use a set of 9 magnitude values ranging from 0.2~mag fainter to 0.2~mag brighter than the initial estimate from \galfitm. We also use a set of 9 $r_e$ values ranging from $3\times$ smaller to $3\times$ larger than the initial estimate. Holding $m_{\textnormal{\scriptsize{F475W}}}$, position angle, and axis ratio fixed, we calculate $\chi^2/\nu$ across this $9\times9$ grid of $m_{\textnormal{\scriptsize{F814W}}}$ and $r_e$ values. We then repeat with $m_{\textnormal{\scriptsize{F814W}}}$ fixed, using a grid of $m_{\textnormal{\scriptsize{F475W}}}$ and $r_e$. We construct contours of $\Delta\chi^2$ in each magnitude--size parameter space, and we identify the 68\% range for $r_e$ based on $\Delta\chi^2=2.3$ \citep{avn76,wal03}. The best-fit values and 68\% limits for $r_e$ are very similar for the $m_{\textnormal{\scriptsize{F814W}}}$--$r_e$ and $m_{\textnormal{\scriptsize{F475W}}}$--$r_e$ grids, and we report the average of the best-fit values and the uncertainty based on the minimum and maximum limits in Table~\ref{tab:re}. To provide additional information about the uncertainty on the $r_e$ estimates and the differences between the F475W and F814W images, we also calculate $\chi^2/\nu$ for each image individually and report the best-fit values in Table~\ref{tab:re}. 

We find half-light radii that are remarkably small ($r_e=0.01\arcsec$--$0.03\arcsec$ for most galaxies), with typical statistical uncertainties of $20\%$--$30\%$. In addition, we find that the half-light radii estimated from the F475W images are systematically smaller than those estimated from the F814W images, by about 40\% on average. This difference between F475W and F814W $r_e$ values is comparable to the width of the 68\% confidence interval on $r_e$ from the multi-band analysis, and it provides an additional diagnostic of the uncertainty for these $r_e$ estimates. As described above (and shown in Figure~\ref{fig:hst_images_1} and Figure~\ref{fig:hst_images_2}), the images at shorter wavelengths have less extended emission, such that the two-dimensional light profiles are more dominated by the central source for the F475W images compared to the F814W images. In other words, our model assumption that the observed light profile can be described by a single component is most appropriate for the F475W images, and the additional extended emission at F814W appears to be driving the model towards larger $r_e$ values. While this suggests that the smaller $r_e$ values estimated from the F475W images might be closest to the ``true" values, we adopt the more conservative estimate of $r_e$ from the joint fit to both UVIS bands, which gives values that are larger than the F475W-only estimate and smaller than the F814W-only estimate \citep[i.e., smaller than F814W-only values from previous work:][]{dia12,sel14}. In Table~\ref{tab:re}, we quantify these differences for each galaxy relative to the fiducial $r_e$ value from a joint fit.   

\begin{deluxetable}{cccc}
\tablecaption{Half-light radius measurements for the central component from \galfitm \label{tab:re}}
\tablehead{
\colhead{} & \colhead{$r_{e,central}$ [$\arcsec$]} & \colhead{} &  \colhead{} \\
\colhead{Galaxy} & \colhead{UVIS} & \colhead{F475W} & \colhead{F814W} }
%\colnumbers
\decimalcolnumbers
\startdata
J0826 & $0.0151\pm0.0031$ & 0.0122 ($-19\%$) & 0.0201 ($+33\%$) \\ % 21, -39 1 0
J0901 & $0.0149\pm0.0033$ & 0.0103 ($-31\%$) & 0.0235 ($+58\%$) \\ % 22, -56 0 0
J0905 & $0.0105\pm0.0027$ & 0.0095 ($-9\%$)  & 0.0117 ($+12\%$) \\ % 26, -19 1 1
J0944 & $0.0099\pm0.0030$ & 0.0072 ($-28\%$) & 0.0137 ($+37\%$) \\ % 30, -47 1 0
J1107 & $0.0156\pm0.0041$ & 0.0112 ($-28\%$) & 0.0238 ($+53\%$) \\ % 26, -53 0 0 
J1219 & $0.0257\pm0.0038$ & 0.0216 ($-16\%$) & 0.0299 ($+16\%$) \\ % 15, -28 0 0 
J1341 & $0.0127\pm0.0023$ & 0.0119 ($-6\%$)  & 0.0138 ($+9\%$)  \\ % 18, -14 1 1
J1506 & $0.0118\pm0.0025$ & 0.0089 ($-24\%$) & 0.0160 ($+36\%$) \\ % 21, -44 0 0 
J1558 & $0.0387\pm0.0064$ & 0.0286 ($-26\%$) & 0.0573 ($+48\%$) \\ % 17, -50 0 0 
J1613 & $0.1289\pm0.0157$ & 0.1072 ($-17\%$) & 0.1604 ($+24\%$) \\ % 12, -33 0 0 
J2116 & $0.0216\pm0.0046$ & 0.0171 ($-21\%$) & 0.0253 ($+17\%$) \\ % 21, -32 1 1
J2140 & $0.0145\pm0.0045$ & 0.0077 ($-47\%$) & 0.0262 ($+81\%$) \\ % 31, -71 1 0
\enddata
\tablecomments{Column 1: Short SDSS name (see Table~\ref{tab:sample}). Column 2: Half-light radius for the central component in arcseconds estimated from a joint fit to the F475W and F814W images. The uncertainty represents a 68\% confidence interval (see Section~\ref{sec:uvis_re}). The median uncertainty is 21\%. Column 3: Best-fit half-light radius based on the F475W images. The quantity in parentheses indicates the percent difference relative to the half-light radius from Column 2. The median difference is $-23$\%. Column 4: Best-fit half-light radius based on the F814W images. The quantity in parentheses indicates the percent difference relative to the half-light radius from Column 2. The median difference is $+35$\%.}
\end{deluxetable}

\subsection{Three-band photometry of the central source}\label{sec:galfitm}

For our three-band photometry of the central source, we use the $r_e$ values measured from the UVIS images as a fixed constraint for simultaneous model fits to the F475W, F814W, and F160W images. For this analysis, we use the $0.05"$/pixel images because this allows us to incorporate the F160W images and it provides the same pixel sampling in all three images. To measure this photometry and uncertainty, which is based on the total flux from model Sersic profiles, we use a method similar to that described above in Section~\ref{sec:uvis_re}. For each magnitude ($m_{\textnormal{\scriptsize{F475W}}}$, $m_{\textnormal{\scriptsize{F814W}}}$, $m_{\textnormal{\scriptsize{F160W}}}$) for each galaxy, we generate \galfitm\ models over a set of 9 magnitude values ranging from 0.2~mag fainter to 0.2~mag brighter than our initial estimate. We then calculate $\chi^2/\nu$ across a $9\times9$ grid of $m_{\textnormal{\scriptsize{F475W}}}$ and $m_{\textnormal{\scriptsize{F814W}}}$ and across a $9\times9$ grid of $m_{\textnormal{\scriptsize{F814W}}}$ and  $m_{\textnormal{\scriptsize{F160W}}}$, and we determine the best-fit magnitude and 68\% range based on contours of $\Delta\chi^2$. We apply corrections for Galactic extinction based on $E(B-V)$ color excess values from the \citet{sch98} infrared dust maps, using the \citet{sch11} calibration for $R_V=3.1$ and the \citet{fit99} reddening law. We report the calibrated photometry of the central component in Table~\ref{tab:nucphot}. As we discuss further in Section~\ref{sec:stellarpop}, the [F475W]$-$[F814W] and [F814W]$-$[F160W] colors are quite blue, implying young ages for the underlying stellar population. 

\begin{deluxetable}{cccc}
\tablecaption{Photometric measurements of the central source from \galfitm \label{tab:nucphot}}
\tablehead{
\colhead{Galaxy} & \colhead{F475W} & \colhead{F814W} & \colhead{F160W} 
}
\decimalcolnumbers
\startdata
J0826 & $19.402\pm0.069$ & $19.050\pm0.061$ & $18.801\pm0.034$ \\%& $0.352\pm0.092$ & $0.249\pm0.070$ \\
J0901 & $19.330\pm0.057$ & $18.944\pm0.065$ & $18.639\pm0.038$ \\%& $0.385\pm0.086$ & $0.306\pm0.075$ \\
J0905 & $19.428\pm0.059$ & $19.065\pm0.061$ & $18.937\pm0.044$ \\%& $0.363\pm0.085$ & $0.128\pm0.075$ \\
J0944 & $19.731\pm0.068$ & $19.165\pm0.070$ & $18.674\pm0.033$ \\%& $0.565\pm0.098$ & $0.491\pm0.077$ \\
J1107 & $19.380\pm0.061$ & $19.095\pm0.071$ & $18.949\pm0.040$ \\%& $0.285\pm0.094$ & $0.146\pm0.081$ \\
J1219 & $19.646\pm0.064$ & $18.823\pm0.061$ & $18.379\pm0.038$ \\%& $0.823\pm0.088$ & $0.443\pm0.072$ \\
J1341 & $18.910\pm0.046$ & $18.756\pm0.056$ & $18.960\pm0.044$ \\%& $0.154\pm0.072$ & $-0.204\pm0.071$ \\
J1506 & $19.074\pm0.050$ & $18.907\pm0.056$ & $18.807\pm0.036$ \\%& $0.167\pm0.075$ & $0.101\pm0.067$ \\
J1558 & $19.611\pm0.064$ & $19.263\pm0.070$ & $18.791\pm0.040$ \\%& $0.347\pm0.095$ & $0.473\pm0.081$ \\
J1613 & $19.451\pm0.073$ & $18.574\pm0.059$ & $17.900\pm0.033$ \\%& $0.877\pm0.094$ & $0.674\pm0.068$ \\
J2116 & $19.591\pm0.089$ & $19.099\pm0.066$ & $19.135\pm0.045$ \\%& $0.491\pm0.111$ & $-0.036\pm0.080$ \\
J2140 & $19.923\pm0.092$ & $19.165\pm0.084$ & $18.992\pm0.044$ \\%& $0.757\pm0.125$ & $0.174\pm0.095$ \\
\enddata
\tablecomments{Column 1: Short SDSS name (see Table~\ref{tab:sample}). Columns 2--4: Calibrated photometry in AB magnitudes for the central component of each galaxy in the F475W, F814W, and F160W images. This is based on a joint-fit to the same morphological component in all three bands, as described in Section~\ref{sec:galfitm}. These values have been corrected for Galactic extinction. The uncertainty represents a 68\% confidence interval.}
\end{deluxetable}

\begin{deluxetable}{ccc}
\tablecaption{Central and total half-light radius \label{tab:re_tot} }
\tablehead{
\colhead{Galaxy} & \colhead{$r_{e,central}$ [kpc]} & \colhead{$r_{e,total}$ [kpc]} }
%\colnumbers
\decimalcolnumbers
\startdata
J0826 & $0.101\pm0.021$ & $0.173^{+0.075}_{-0.053}$  \\
J0901 & $0.087\pm0.019$ & $0.237^{+0.144}_{-0.088}$ \\
J0905 & $0.076\pm0.019$ & $0.097^{+0.044}_{-0.033}$  \\
J0944 & $0.061\pm0.019$ & $0.114^{+0.067}_{-0.047}$ \\
J1107 & $0.092\pm0.024$ & $0.273^{+0.192}_{-0.112}$ \\
J1219 & $0.148\pm0.022$ & $0.412^{+0.194}_{-0.124}$ \\
J1341 & $0.088\pm0.016$ & $0.117^{+0.040}_{-0.032}$  \\
J1506 & $0.079\pm0.017$ & $0.168^{+0.076}_{-0.054}$ \\
J1558 & $0.209\pm0.034$ & $0.778^{+0.383}_{-0.244}$ \\
J1613 & $0.741\pm0.090$ & $0.949^{+0.274}_{-0.207}$ \\
J2116 & $0.157\pm0.033$ & $0.284^{+0.131}_{-0.092}$ \\
J2140 & $0.107\pm0.033$ & $0.153^{+0.092}_{-0.064}$ \\
\enddata
\tablecomments{Column 1: Short SDSS name (see Table~\ref{tab:sample}). Column 2: Half-light radius for the central component in kpc estimated from a joint fit to the F475W and F814W images (see Section~\ref{sec:uvis_re} and Table~\ref{tab:re}). The uncertainty represents a 68\% confidence interval, and the median fractional uncertainty is 21\%. Column 3: Estimate of half-light radius for the entire galaxy based on the F814W images. This involves extrapolating the $n=4$ Sersic model until it reaches 50\% of the total light from the galaxy. The upper and lower limits account for the uncertainty on this extrapolation, including uncertainty on the central $r_e$ value, the central flux, and the total flux. The median difference for the lower limit is $-32\%$ and the median difference for the upper limit is $+47\%$.}
\end{deluxetable}

\subsection{Estimating half-light radius for the entire galaxy}\label{sec:re_total}

The analysis described above provides measurements of the half-light radius and three-band photometry for the central component of each galaxy, which typically contributes $\approx75\%$ of the total light at F814W and $\approx90\%$ of the total light at F475W. Because there is significant residual emission, the half-light radius for the central component contains less than half of the light for the entire galaxy. To provide an estimate of $r_e$ that is more appropriate for the total light from each galaxy, we extrapolate the $n=4$ Sersic model until it reaches half of the total flux measured using the F814W images in circular apertures with 25~kpc diameters. We account for uncertainties on this extrapolation based on the flux uncertainty for the central and total components, which impacts how far one needs to extrapolate the central light profile. In practice, this estimate of total half-light radius is larger than the central half-light radius by a factor of two on average, which is similar to the ratio $r_{70}/r_{50}=2.08$ for an $n=4$ Sersic model. We report these values in Table~\ref{tab:re_tot}. For the remainder of the paper, we use $r_{e,central}$ to refer to the half-light radius of the central component and $r_{e,total}$ to refer to this larger estimate of total half-light radius.

\begin{figure*}[!t]
\begin{center}
\includegraphics[angle=0,scale=1.1]{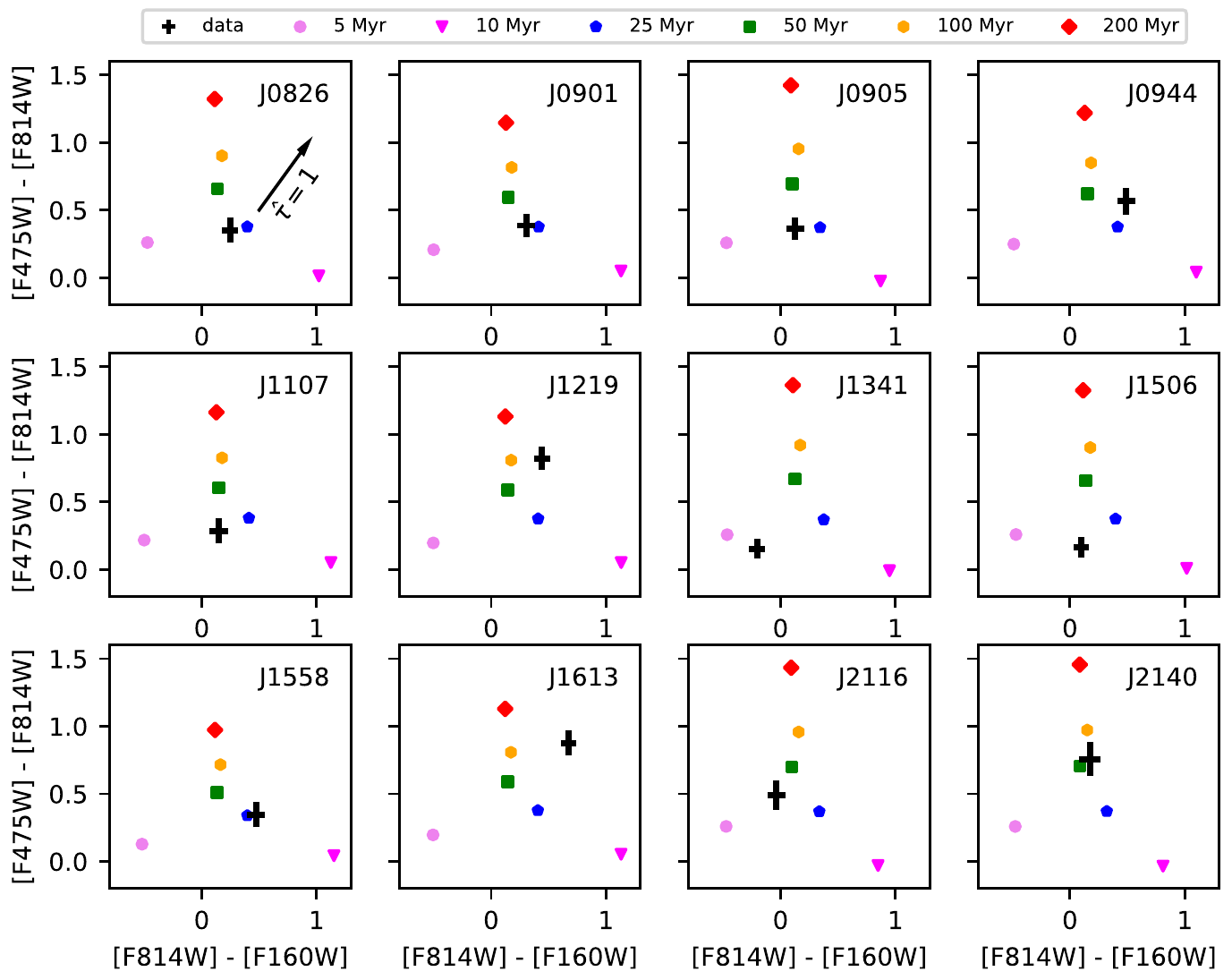}
\caption{Plots of observed-frame [F475W]$-$[F814W] and [F814W]$-$[F160W] colors for each of the 12 galaxies in the sample. Each panel shows the observed colors for one galaxy and for SSP models at the galaxy's redshift with ages between $t=5$~Myr and $t=200$~Myr. The arrow in the top-left panel shows the impact of dust attenuation with optical depth $\hat{\tau}=1$ at $\lrest=5500$~\AA, which corresponds to $A_V=1.086$~mag. While many galaxies are consistent with ages between $t=25$~Myr and $t=50$~Myr, the bluest galaxies (e.g., J1341, J1506) clearly require $t<10$~Myr.}
\label{fig:color_color}
\end{center}
\end{figure*}

\section{Stellar population modeling}\label{sec:stellarpop}

The multi-band photometry of the compact stellar population in each galaxy allows us to estimate the stellar mass associated with these recently formed stars. For comparison, the spatially integrated (i.e., unresolved) photometry of these galaxies, as described in Section~\ref{sec:extended} and used in previous studies \citep[e.g.,][]{dia12,rup19}, includes two ultraviolet bands from the Galaxy Evolution Explorer \citep[GALEX,][far-ultraviolet FUV and near-ultraviolet NUV]{mar05_galex,mor07}, five optical bands from the SDSS (ugriz), and four infrared bands from the Wide-field Infrared Survey Explorer \citep[WISE,][W1, W2, W3, W4]{wri10}. In contrast, the spatially resolved photometry from HST is limited to three bands. This presents a challenge for performing detailed stellar population synthesis modeling with a large range of free parameters that describe the star-formation history and dust properties \citep[e.g.,][]{con13}. 

In this context, we begin by considering simple stellar population (SSP) models from \fsps\ \citep[Flexible Stellar Population Synthesis:][]{con09,con10} to gain insight on parameters such as stellar mass and stellar age. Given that we are primarily interested in determining the properties of the stars that formed within the central several hundred parsecs during a recent starburst event, this idealized assumption of a single-age central stellar population provides useful constraints. To visualize the range of SSP ages that are consistent with our nuclear photometry, we use \fsps\ to compute the expected [F475W]$-$[F814W] and [F814W]$-$[F160W] observed-frame colors for a grid of models with ages between $t=5$~Myr and $t=200$~Myr at the redshift of each of our 12 galaxies. This comparison avoids the need to make $K$-corrections \citep[e.g.,][]{hog02,bla07} to compute magnitudes and colors in a common set of rest-frame filters.  

As shown in Figure~\ref{fig:color_color}, we find that all 12 galaxies are consistent with light-weighted stellar ages for their central starburst components in the range $t=5$--50~Myr. At least two galaxies (J1341, J1506) clearly require a central stellar age $t<10$~Myr, and a few other galaxies (J0905, J1107, J2116) have blue colors that also suggest very young stellar ages (i.e., implying a starburst event within the last 10~Myr). A number of galaxies (J0826, J0901, J0944, J1558) have colors that are consistent with a $t\approx25$~Myr stellar population with small dust attenuation, and the remaining galaxies with the reddest [F475W]$-$[F814W] colors (J1219, J1613, J2140) are consistent with more dust attenuation and/or older ages $t\approx50$~Myr. As a preliminary estimate, we compute the stellar mass associated with a $t=25$~Myr SSP at solar metallicity, scaled to match the F814W observed-frame luminosity for each galaxy. This yields a median value of $\log(\mstar{}_{,central})=10.3$, which matches the median value we find from our more detailed analysis below (see Section~\ref{sec:nucmass}). For reference, forming this amount of stellar mass over a timescale of 25~Myr implies an average star-formation rate of $\approx800~\msun/\textnormal{yr}^{-1}$. This value is large, but it is within the range of previously published star-formation rates for this sample based on infrared luminosity and SED modeling \citep[e.g.,][]{dia12,gea13,sel14,gea14,gea18}.       

\begin{figure*}[!t]
\begin{center}
\includegraphics[angle=0,scale=0.67]{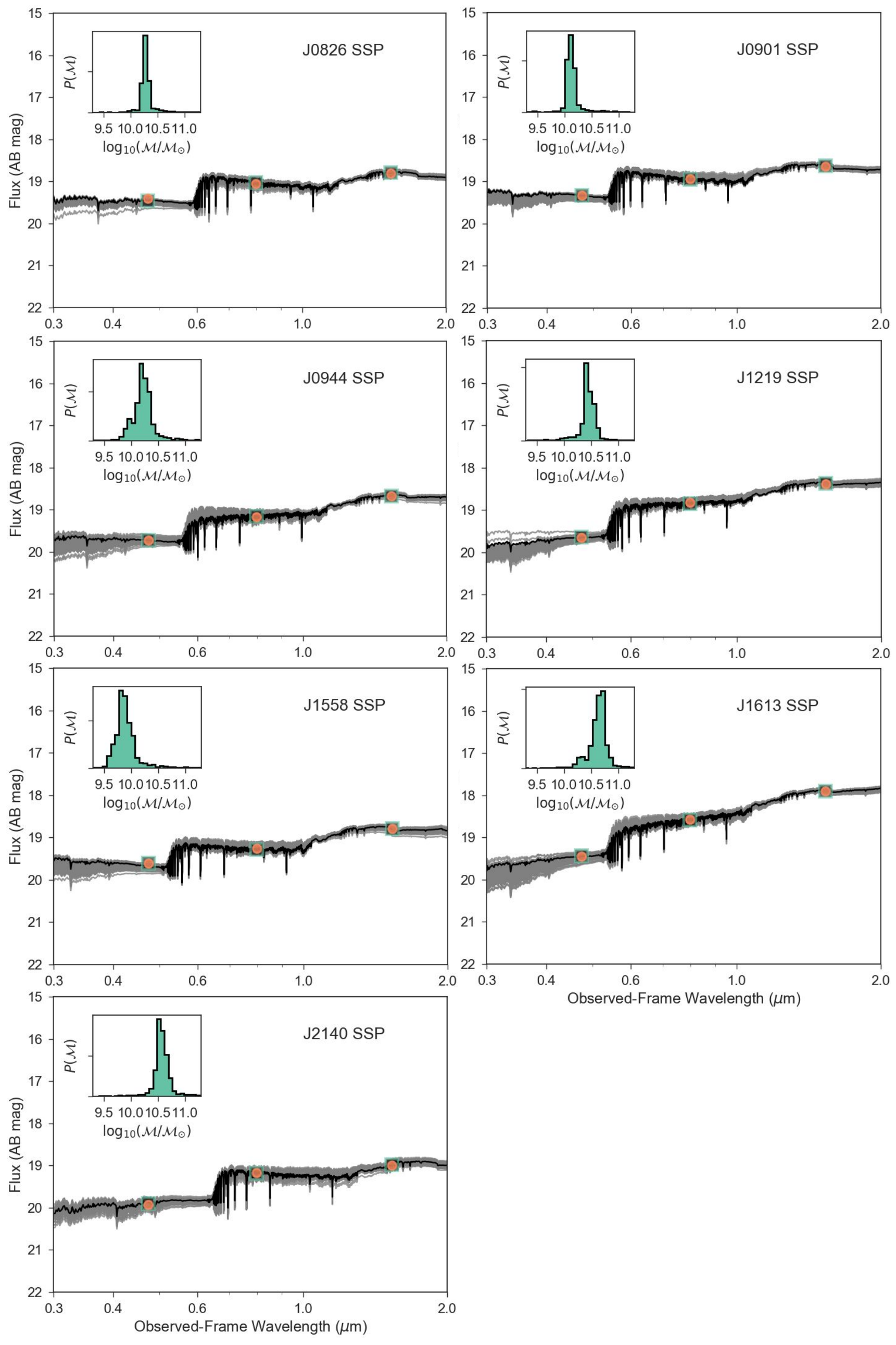}
\end{center}
\caption{Results from SSP model fits to the nuclear SEDs for the 7/12 galaxies that are consistent with ages $t\geq25$~Myr for their central component. The main panels show the observed nuclear SEDs (orange circles), a random sampling ($n=100$) of the SSP models considered by \prospector\ (gray shading), the maximum-likelihood SSP model (black line), and model photometry for the maximum-likelihood model (green boxes). The inset panels show the stellar mass posteriors from \prospector. In all seven cases, the SSP models provide reasonable fits to the photometry and the stellar mass is well constrained.}
\label{fig:ssp_sed}
\end{figure*}

\begin{figure*}[!th]
\begin{center}
\includegraphics[angle=0,scale=0.696]{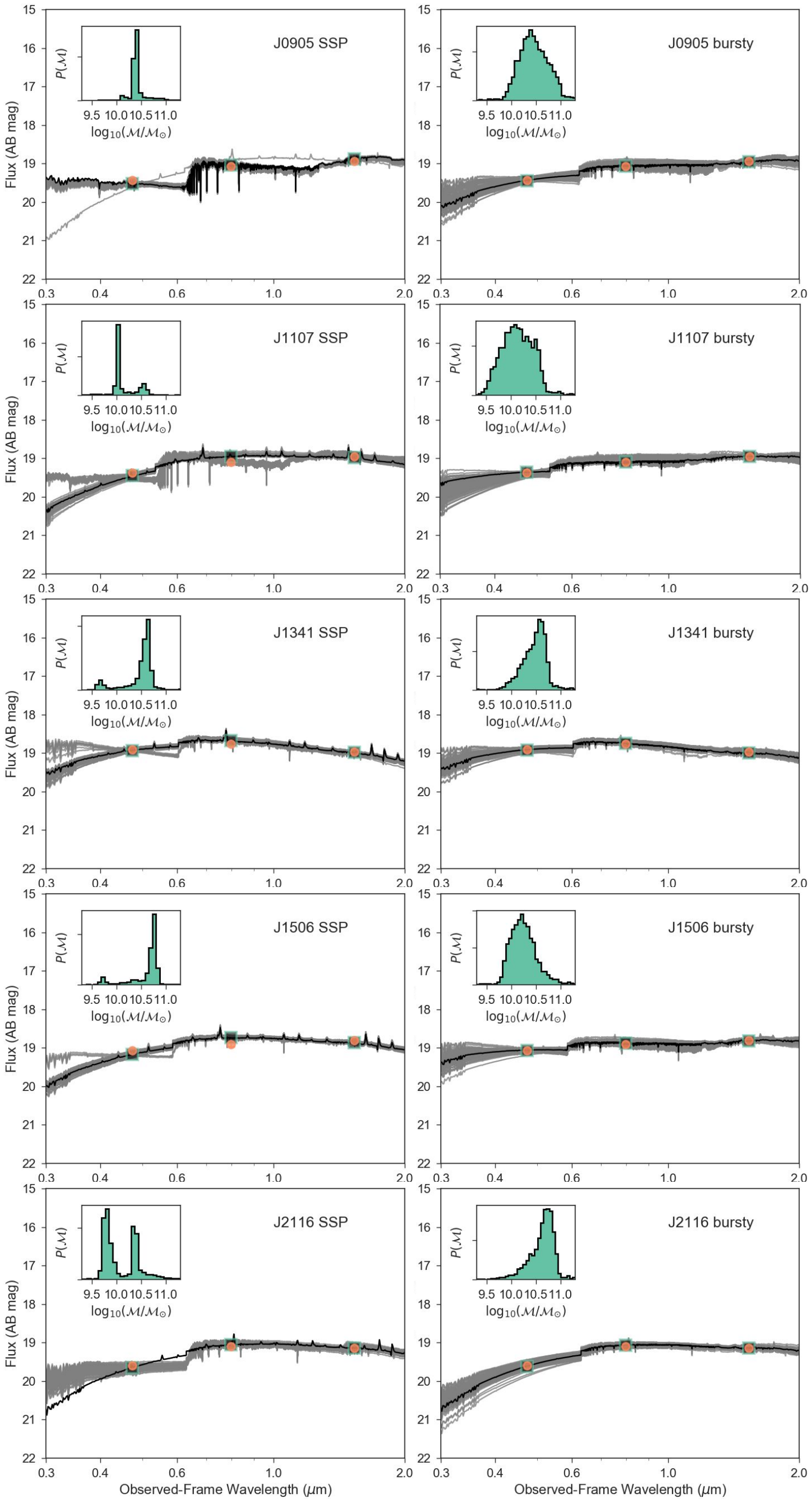}
\caption{Results from stellar population modeling for the 5/12 galaxies with the bluest colors, which suggest significant contributions from stars with $t\leq10$~Myr. Each panel shows an observed nuclear SED (orange circles), a random sampling ($n=100$) of the models considered by \prospector\ (gray shading), the maximum-likelihood model (black line), and model photometry for the maximum-likelihood model (green boxes), while the inset panel show the stellar mass posterior. Left: Results from SSP models, for which the model fits and the posterior distributions show degeneracies. The models with stronger Balmer breaks are older, less dusty, and less massive, while the models with weaker Balmer breaks are younger, more dusty, and more massive. Right: The same SEDs with model fits for a more flexible star-formation history: a delayed-$\tau$ model with a late-time burst. These models more robustly characterize the range of potential SEDs and the uncertainty on the stellar mass.}
\label{fig:bursty_sed}
\end{center}
\end{figure*}

\subsection{Probability distributions for central stellar mass}\label{sec:nucmass}

With the goal of quantifying posterior probability distributions for stellar mass, we use \prospector\ \citep{lej17,joh19} with model assumptions and prior probabilities informed by the SSP comparisons above. Within \prospector, we use the \dynesty\ package \citep{spe20}, which estimates Bayesian posteriors using dynamic nested sampling. We begin with SSP models ($\textnormal{sfh}=0$ in \fsps) fixed at solar metallicity, and we adopt a prior probability on age that is flat in logarithmic spacing between 3~Myr and 100~Myr, informed by the blue colors and young ages found above. We also implement dust attenuation as a power law ($\textnormal{dust\_type}=0$ in \fsps) with the optical depth $\hat{\tau}\propto\lambda^{-0.7}$ \citep{cha00}. 

We find that the galaxies with less extreme colors (i.e., those that are consistent with $t\geq25$~Myr) have nuclear SEDs that can be well described by single-age SSP models. In Figure~\ref{fig:ssp_sed}, we show the observed photometry, model photometry, model SEDs, and stellar mass posterior distributions for the galaxies J0826, J0901, J0944, J1219, J1558, J1613, and J2140. The width of the stellar mass posterior between the 16th and 84th percentiles (i.e., corresponding to a 68\% confidence interval) ranges from 0.09~dex to 0.25~dex for these seven galaxies. The posterior distributions are also single-peaked and reasonably symmetrical, indicating that the stellar mass value is well constrained. 

In contrast, the nuclear SEDs of galaxies with the bluest colors (consistent with $t<10$~Myr) are not fit robustly by single-age SSP models. In particular, there is degeneracy between SSP model ages that is evident in the range of SSP model fits. We show SSP model fits on the left-hand side of Figure~\ref{fig:bursty_sed}, and we find comparable fits to the photometry for models that are different in age by almost an order of magnitude (e.g., $t\approx4$~Myr vs $t\approx30$~Myr for J2116). For all five galaxies, this degeneracy is between models with stronger Balmer breaks (which are older, less dusty, and less massive) compared to models with weaker Balmer breaks (which are younger, more dusty, and more massive). This degeneracy is also evident in the asymmetries and multiple peaks in the stellar mass posteriors (e.g., the width of the stellar mass posterior between the 16th and 84th percentiles is larger than 0.5~dex for galaxies J1107 and J2116). 

To provide more robust constraints on the central stellar mass for these younger and bluer galaxies, we explore models with more flexible star-formation histories. In particular, we use delayed-$\tau$ models ($\textnormal{sfh}=4$ in \fsps) for the star formation history ($\textnormal{SFR}\propto t e^{-t/\tau}$) with additional parameters to characterize a recent burst of star formation, including the time at which the burst happens and the fraction of the total stellar mass formed in the burst. This approach is very similar to the stellar population modeling described by \citet{rup19}. In the right panel of Figure~\ref{fig:bursty_sed}, we show that these models are clearly capable of fitting the photometry. The stellar mass posterior distributions are often broader than those for SSP models, reflecting how the more flexible models have a broader range of parameters that are consistent with the photometry. Compared to results from SSP modeling, this provides constraints on stellar mass that are less sensitive to uncertainties in the star-formation history. The difference between the median of the stellar mass posterior and the 16th and 84th percentile values is $\pm0.2$~dex or $\pm0.3$~dex for all five galaxies. In comparing the posterior median values between the SSP models and the delayed-$\tau$ models with bursts, the values agree within 0.1~dex for J0905, J1107, and J1341. In contrast, there are large differences for J1506 (0.6~dex) and J2116 (0.8~dex), both of which have more robust constraints from the more flexible models.

\begin{figure}[!h]
\begin{center}
\includegraphics[angle=0,scale=0.6]{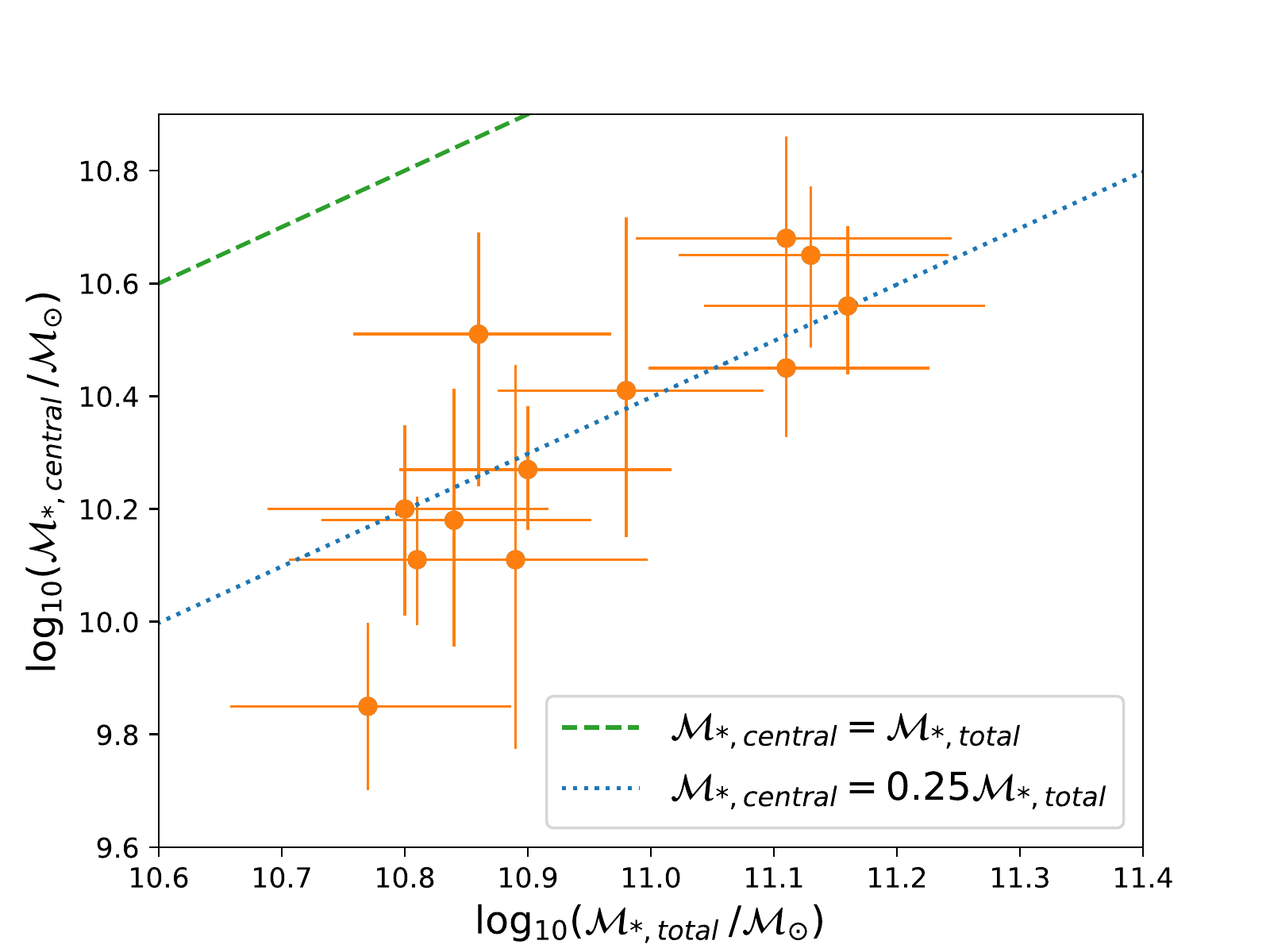}
\caption{The relationship between central stellar mass and total stellar mass. The central stellar mass is estimated from spatially resolved nuclear SEDs, and the total stellar masses is estimated from spatially unresolved total SEDs. The vast majority of the sample is consistent with $\mstar{}_{,central}/\mstar{}_{,total}\approx0.25$.}
\label{fig:central_total}
\end{center}
\end{figure}

\begin{figure}[!h]
\begin{center}
\includegraphics[angle=0,scale=0.6]{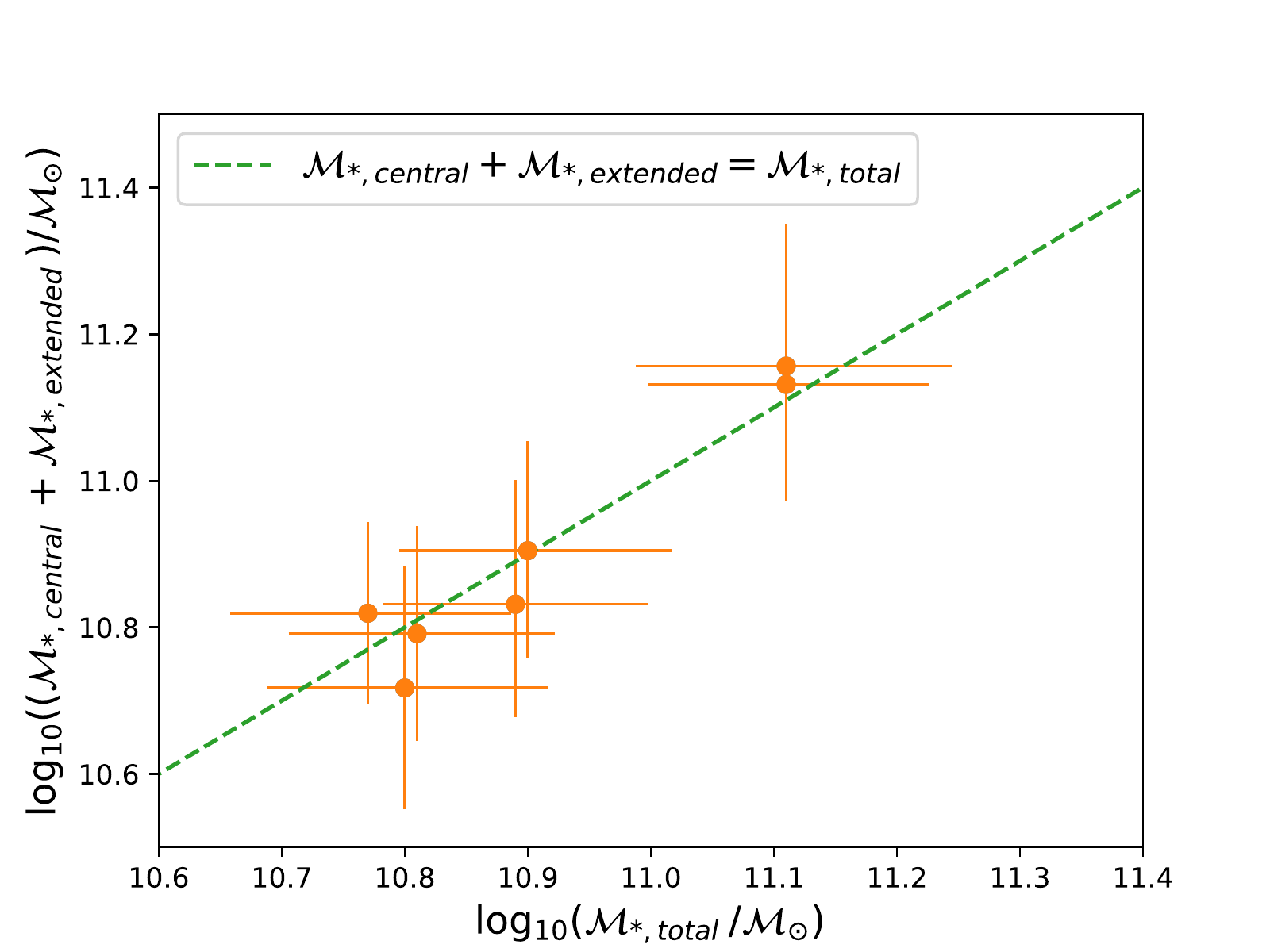}
\caption{A comparison between measurements of central stellar mass $+$ extended stellar mass from spatially resolved SEDs (using HST data) and independent measurements of total stellar mass from spatially unresolved SEDs (not using HST data). This focuses on the seven galaxies in Table~\ref{tab:resid} with the smallest uncertainties on F475W extended flux. These two methods for estimating the total stellar mass agree well within the uncertainties. }
\label{fig:nuc_ext_tot}
\end{center}
\end{figure}

\subsection{Comparing to the Total Stellar Population}\label{sec:extended}

Our analysis of the central stellar population can be combined with analysis of the total stellar population to determine the fraction of the total stellar mass associated with the compact starburst component. To estimate the total stellar mass, we make use of publicly available photometric measurements at ultraviolet (GALEX Release 7), optical \citep[SDSS Data Release 15,][]{agu15}, and infrared \citep[unWISE,][]{lan14,lan16} wavelengths. We use the modelMag photometry from SDSS, calibrated magnitudes from GALEX, and the unWISE forced photometry based on SDSS positions. For galaxies without GALEX fuv\_mag values (J0944, J1219, J2116), we use GALEX FUV fluxes measured at the NUV positions. We focus on the WISE W1 and W2 bands (central wavelengths 3.4 and 4.6~$\mu$m) because these wavelengths are dominated by stellar emission, and we do not include the W3 and W4 bands (central wavelengths 12 and 22~$\mu$m) in our stellar population analysis because these wavelengths are dominated by dust emission. The total SEDs span an observed wavelength range 0.15--4.6~$\mu$m (corresponding to $\lrest=0.10$--2.9~$\mu$m at the median redshift of our sample). We apply corrections for Galactic extinction in each band based on extinction coefficients from \citet{yua03} for the \citet{fit99} reddening law with 14\% corrections to $E(B-V)$ values from \citet{sch98}.  

For the stellar population modeling with \prospector, we use a delayed-$\tau$ model with a late-time burst for the star-formation history. We impose a lower limit on stellar age (i.e., how long ago the galaxy began forming stars) of 1~Gyr and an upper limit of 7~Gyr, which is approximately the age of the universe at $z=0.75$, the highest redshift in the sample. We also allow up to 50\% of the stars to have formed in the recent burst of star formation. We report the constraints on stellar mass ($\mstar$), stellar age ($t_{age}$), the e-folding parameter ($\tau$), the burst fraction ($f_{burst}$), and the time of the burst expressed as a fraction of the stellar age ($f_{age,burst}$) in Table~\ref{tab:tot_params}. For the total stellar mass posteriors, we find a median for the 50th percentile values across the sample of $\log(\mstar/\msun)=10.9$. In comparison, the median for the central stellar mass posteriors was $\log(\mstar/\msun)=10.3$. We show the relationship between central stellar mass and total stellar mass in Figure~\ref{fig:central_total}. The vast majority of the galaxies (11/12) are consistent with $\mstar{}_{,central}/\mstar{}_{,total}\approx0.25$. The only exception is J1558 ($\mstar{}_{,central}/\mstar{}_{,total}\approx0.12$), which is an ongoing merger (see Section~\ref{sec:sample} and Figure~\ref{fig:hst_images_2}) that is the least dominated by a single source (e.g., it has a lot of extended emission, including a second, fainter nucleus). 

As a consistency check, we also use the HST images to measure the fluxes and colors of the extended, non-nuclear emission in each galaxy with the goal of measuring the extended stellar mass. We do this by subtracting the central Sersic photometry from the total galaxy photometry measured in apertures with 25~kpc diameters. In practice, the uncertainties on these extended fluxes can be large, especially in the UVIS bands, because the median ratio of central flux to total flux is 93\% at F475W, 75\% at F814W, and 66\% at F160W. The typical $\approx6\%$ uncertainty on the central fluxes in the UVIS bands then propagates into a much larger fractional uncertainty for the fainter extended flux. Focusing on the seven galaxies with the smallest uncertainties on their extended F475W flux, we carry out a stellar population analysis based on extended emission in the F475W, F814W, and F160W images. We use delayed-$\tau$ star-formation histories with parameters to characterize a more recent burst, and we show the constraints on stellar population parameters in Table~\ref{tab:resid}. We compare $\mstar{}_{,central} + \mstar{}_{,extended}$ estimated from the spatially resolved HST photometry to $\mstar{}_{,total}$ estimated from spatially integrated photometry in Figure~\ref{fig:nuc_ext_tot}, and these two methods of estimating the total stellar mass show very good agreement. 

\begin{figure*}[!t]
\begin{center}
\includegraphics[angle=0,scale=1.1]{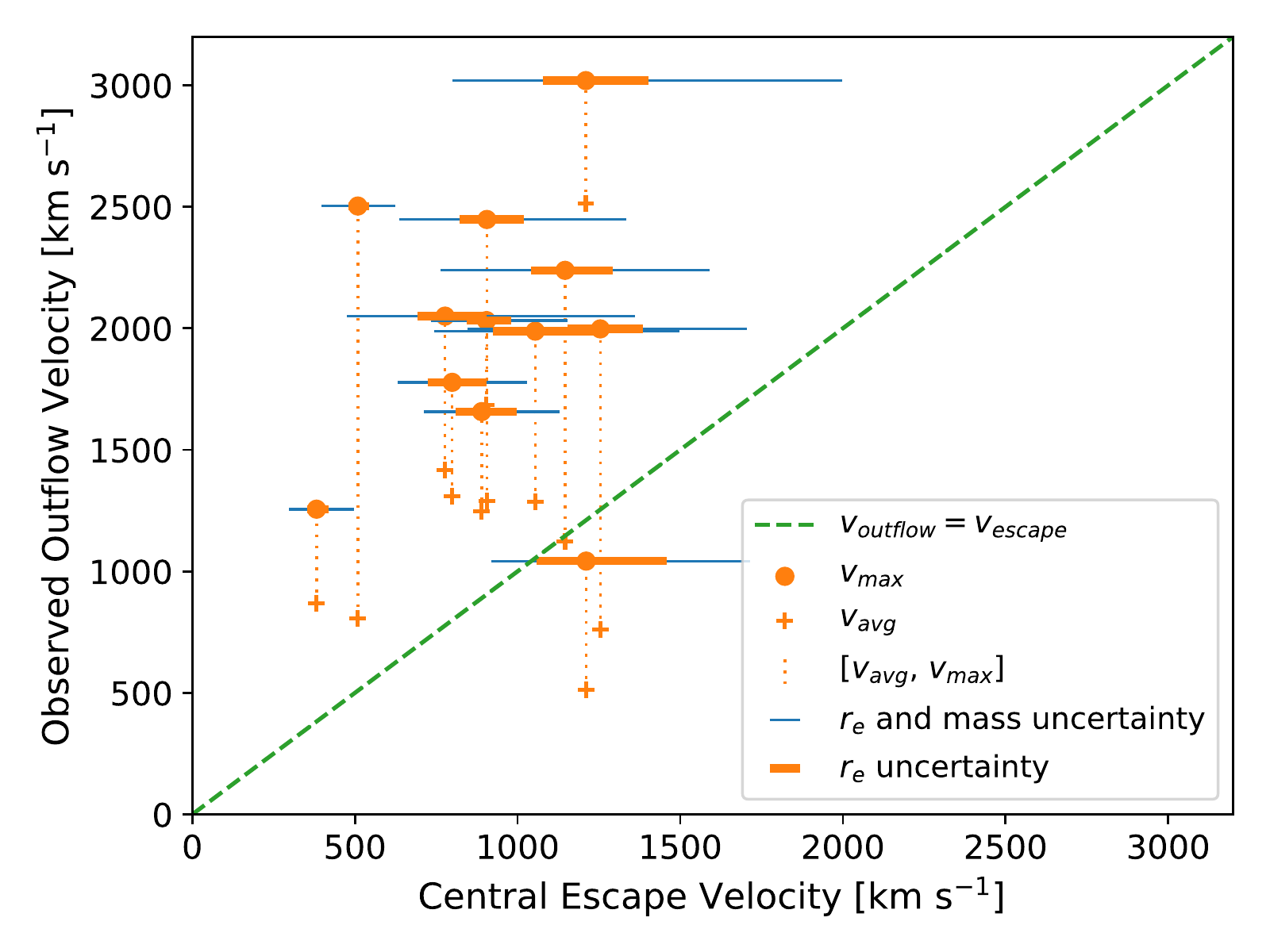}
\caption{A comparison between observed outflow velocity and central escape velocity. The average outflow velocity is based on the median of the cumulative equivalent width distribution for the $\mgii~\lambda\lambda2796,2803$ doublet (see Table~\ref{tab:sample}); the maximum outflow velocity is based on the point where the equivalent width distribution for \mgii~$\lambda2796$ reaches 95\% of the total (see Table~\ref{tab:sample}); and the central escape velocity is based on central stellar mass and effective radius (see Equation~\ref{eqn:esc}). The maximum outflow velocity exceeds the central escape velocity for 11/12 galaxies, and the median ratio is $v_{max} / v_{esc,central}=2.2$. The average outflow velocity exceeds the central escape velocity for 9/12 galaxies, and the median ratio is $v_{avg} / v_{esc,central}=1.5$. This significant excess of outflow velocity relative to escape velocity has important implications for the physical mechanisms responsible for driving the fast outflows and for the fate of the outflowing gas (see Section~\ref{sec:escape_surface}).}
\label{fig:vesc}
\end{center}
\end{figure*}

\section{Discussion}\label{sec:escape_surface}

The primary motivation for obtaining these multi-band imaging data was to test physical mechanisms that could launch outflows (1) at speeds significantly faster than the central escape velocity or (2) at speeds comparable to the central escape velocity. In \citet{dia12}, we found evidence for star-formation rate surface densities that approach the Eddington limit \citep[$\sigmasfr\sim3000$~\units;][]{mur05,tho05,hop10} and we pointed out that if the stellar mass profile matched the observed light profile, then it would be possible to have escape velocities that were comparable to the observed outflow velocities. This would have been consistent with scenario (2) and models of radiation-driven winds that predict outflow velocities near the escape velocities of the densest stellar systems within a galaxy \citep[e.g.,][]{mur11}. In contrast, our new results (see below and Figure~\ref{fig:vesc}) are consistent with scenario (1) and models that can produce $>1000$~\kms\ outflows that exceed the central escape velocity by a significant factor \citep[e.g.,][]{hec11,tho15,bus16}. In other words, the fact that the compact starburst component contributes 25\% to the total stellar mass implies that 10\%--15\% of the total stellar mass is within the half-light radius (i.e., half the mass of the compact starburst is within the half-light radius). This yields values for central escape velocity (Section~\ref{sec:vesc}) and central stellar surface density (Section~\ref{sec:sigma}) that require scenario (1) to explain the fast outflows. We explore this comparison with outflow models in more detail in Section~\ref{sec:fast}. 

\subsection{Constraints on Central Escape Velocity}\label{sec:vesc}

We can combine the measurements of half-light radius (Table~\ref{tab:re}, Section~\ref{sec:uvis_re}) with the measurements of central stellar mass (Table~\ref{tab:sspmass}, Table~\ref{tab:burstmass}, Section~\ref{sec:nucmass}) to estimate a central escape velocity for each galaxy. 
\begin{equation}\label{eqn:esc}
v_{esc,central} = \sqrt{G \mstar{}_{,central} / r_{e,central}}
\end{equation}
This form of Equation~\ref{eqn:esc} uses the entire stellar mass of the central stellar population, and it assumes that half of that mass is within the half-light radius. Based on this calculation, we find a median value of $v_{esc,central}=900$~\kms, which is very similar to the estimated central escape velocity of the Milky Way on $r=0.1$~kpc scales \citep[e.g.,][]{ken08,bro15}.  

We compare these estimates of central escape velocity with observed outflow velocity in Figure~\ref{fig:vesc}. We use the lower limits on $r_{e,central}$ and upper limits on $\mstar{}_{,central}$ to calculate upper limits on $v_{esc,central}$, and we use the opposite limits on size and mass to calculate lower limits on $v_{esc,central}$. We also show the portion of the uncertainty on $v_{esc,central}$ and that comes from the upper and lower limits on $r_{e,central}$. When characterizing the outflow velocity by $v_{max}$ (defined by the point at which the equivalent width distribution for $\mgii~\lambda2796$ reaches 95\% of the total, see Table~\ref{tab:sample}) we find that 11/12 galaxies have observed outflow velocities that clearly exceed their central escape velocities, with a median $v_{max} / v_{esc,central}=2.2$ for the full sample. The only galaxy consistent with $v_{max}=v_{esc,central}$ is J2140, which has the slowest outflow speed in the sample and is also the only galaxy with evidence for a non-zero AGN continuum contribution at near-ultraviolet wavelengths (see Section~\ref{sec:sample}). This contribution could bias the $r_{e,central}$ estimate to be low for J2140 (especially when considering the F475W band, see Table~\ref{tab:re}) and the $v_{esc,central}$ estimate to be high. If we used the $r_{e,central}$ value based solely on F814W, this galaxy would have a best-fit ratio $v_{max}/v_{esc,central}=1.15$, but it would still be consistent with $v_{max}=v_{esc,central}$ given the uncertainties. Regardless, this does not impact our estimate of the median $v_{max} / v_{esc,central}$ ratio for the sample. 

We also find that the average outflow velocity $v_{avg}$ (defined by the median of the cumulative equivalent width distribution for $\mgii~\lambda\lambda2796,2803$, see Table~\ref{tab:sample}) exceeds the best-fit central escape velocity for 9/12 galaxies, with $v_{avg} / v_{esc,central}=1.5$ as the median ratio for the full sample. This implies that the bulk of the outflowing gas in the vast majority of these galaxies has sufficient momentum to easily escape from the central regions of the galaxy. Given the typical outflow velocities $\overline{v_{max}}=2000$~\kms and $\overline{v_{avg}}=1200$~\kms\ for this sample, and the fact that several of these galaxies are known to have extended molecular outflows on several kpc scales \citep{gea18} to 10~kpc scales \citep{gea14}, these ionized outflows are also clearly capable of escaping the galaxy into the circumgalactic medium \citep[e.g., the $r=10$--50~kpc ionized outflows found by ][]{rup19}. Unless forces besides gravity are able to slow them down, they would also escape their $10^{12-13}~\msun$ dark matter halos \citep[e.g.,][]{mos13,mur15}. 

\begin{figure*}[!t]
\begin{center}
\includegraphics[angle=0,scale=1.0]{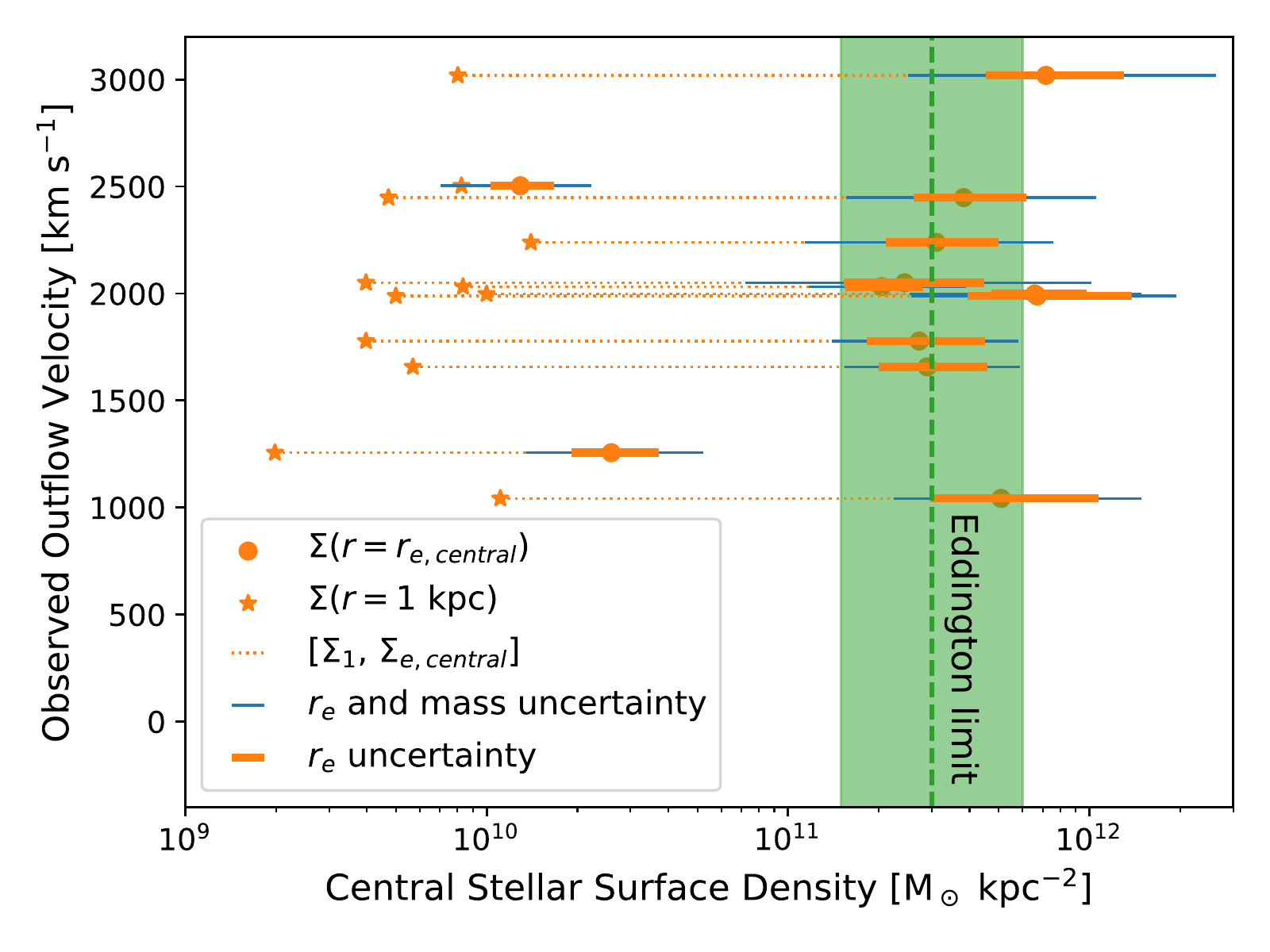}
\caption{A comparison between observed outflow velocity and central stellar surface density. This is based on $v_{max}$ outflow velocities (Table~\ref{tab:sample}), $\Sigma_{e,central}$ surface densities within the half-light radius (Equation~\ref{eqn:sigma}), and $\Sigma_1$ surface densities within the central kpc (calculated as lower limits using Equation~\ref{eqn:sig1}). There is no apparent relationship between outflow velocity and stellar surface density in our sample. The median value $\Sigma_{e,central}=3\times10^{11}$~\sigunits\ on $r=0.1$--0.2~kpc scales matches theoretical estimates of the Eddington limit \citep[e.g.,][]{hop10}, while the median value $\Sigma_{1}=7\times10^{9}$~\sigunits\ on $r=1$~kpc scales is consistent with values for compact galaxies at $0.5<z<3.0$ \citep[e.g.,][]{bar17}. }
\label{fig:sigma_star}
\end{center}
\end{figure*}

\begin{figure*}[!t]
\begin{center}
\includegraphics[angle=0,scale=1.0]{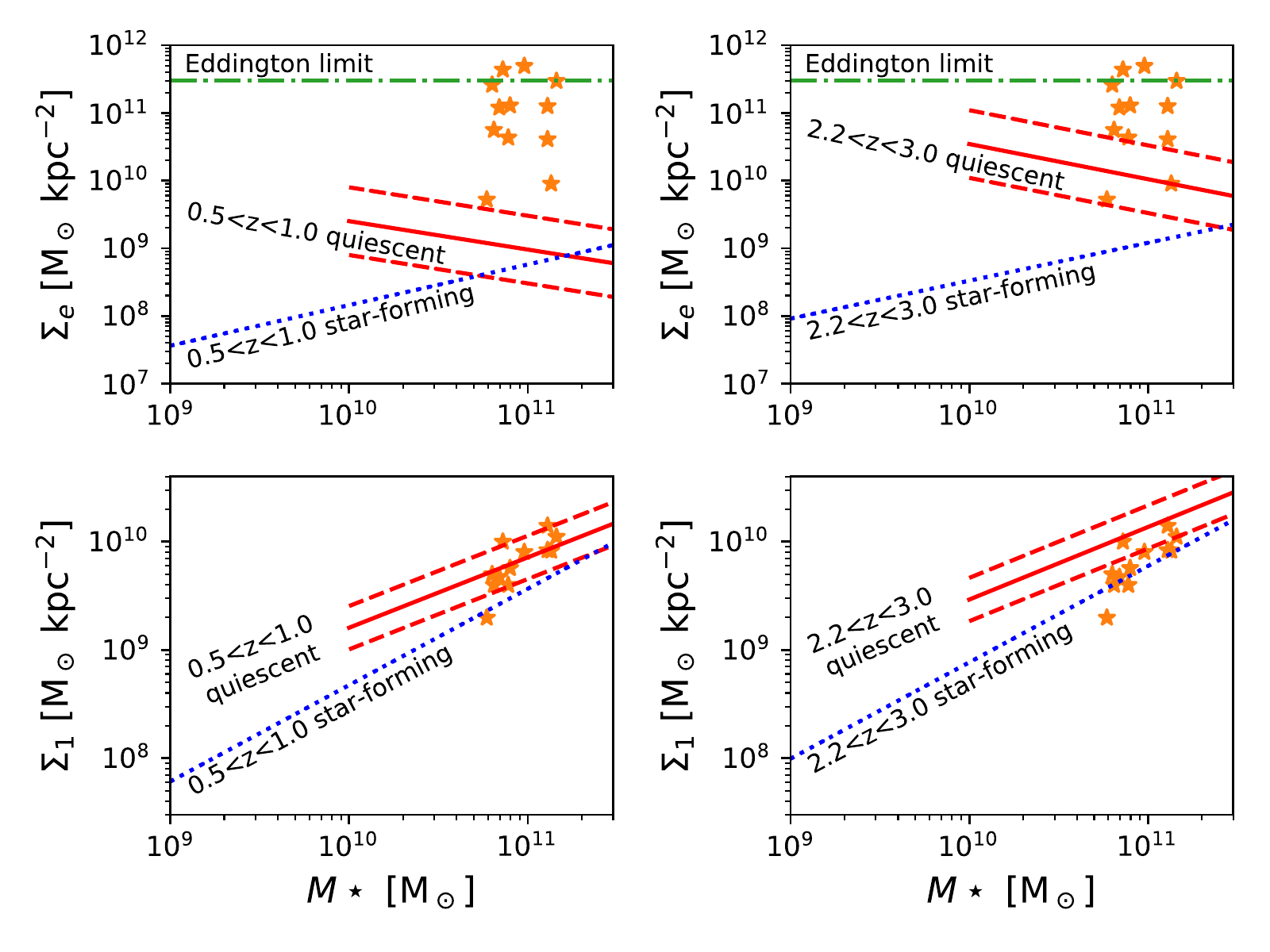}
\caption{The relationship between stellar mass surface density and stellar mass. The points for our sample are based $\Sigma_{e,total}$ values on $r=0.2$--0.3~kpc scales (Equation~\ref{eqn:sig_tot}), $\Sigma_{1}$ values on $r=1$~kpc scales (calculated as lower limits using Equation~\ref{eqn:sig1}), and total stellar mass values (see Table~\ref{tab:sample} and Table~\ref{tab:tot_params}). The solid red and dotted blue lines illustrate the relations for quiescent and star-forming galaxies at $0.5<z<3.0$ from \citet{bar17}. The top panels show $\Sigma_e$, the bottom panels show $\Sigma_1$, the left panels show relations at $0.5<z<1.0$ (overlapping with our sample), and the right panels show relations at $2.2<z<3.0$ (the highest redshift available). In the top panels, the green dash-dot line corresponds to the theoretical estimate of the Eddington limit from \citet{hop10}. The data for the galaxies in our sample are the same for both the left and right panels. In terms of $\Sigma_e$, 10/12 galaxies in our sample (see Section~\ref{sec:z2} for details) lie significantly above the relations found by \citet{bar17} at all redshifts. In terms of $\Sigma_1$, there is good agreement between our lower-limit values and the relation for $0.5<z<1.0$ quiescent galaxies, although several of our galaxies extend slightly above the scatter at $0.5<z<1.0$ and are closer to the relation for $2.2<z<3.0$ quiescent galaxies. This shows that our galaxies are extreme in terms of $\Sigma_e$, but their more modest $\Sigma_1$ values are consistent with those of compact massive galaxies from the literature.}
\label{fig:sigma_mass}
\end{center}
\end{figure*}

\begin{figure*}[!t]
\begin{center}
\includegraphics[angle=0,scale=1.0]{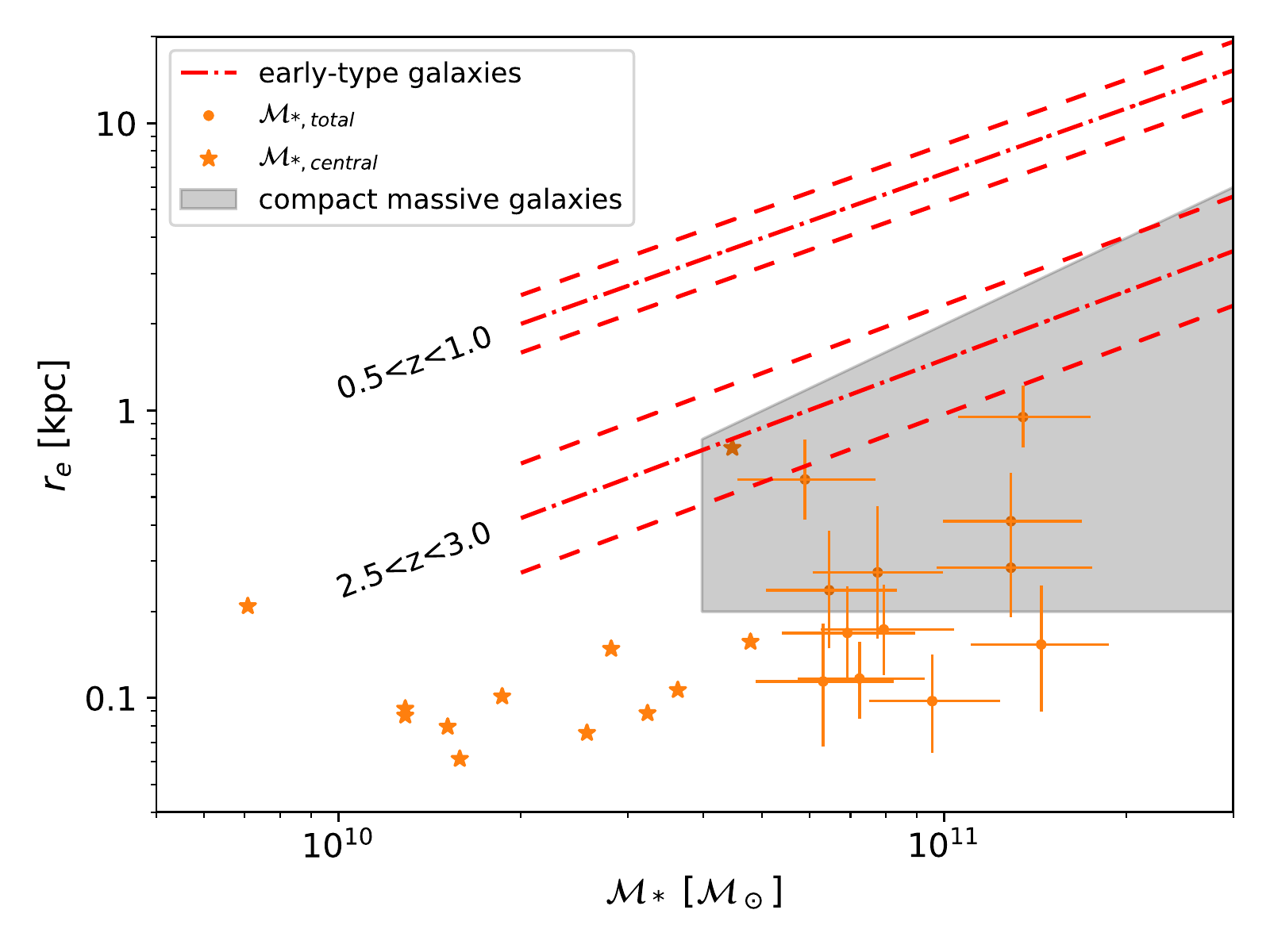}
\caption{The relationship between half-light radius and stellar mass. For our sample, the values for $r_{e,total}$ and total stellar mass are shown as the circles with error bars, and the values for $r_{e,central}$ and central stellar mass are shown as stars. The size--mass relations for early-type galaxies at $0.5<z<1.0$ (overlapping with our sample) and $2.5<z<3.0$ (the highest redshift available) from \citet{van14} are indicated by the red dot-dashed lines. The definition of compact massive galaxies at $2.0<z<2.5$ from \citet{van15} is represented by the gray shaded region. Based on the values for total stellar mass, 10/12 galaxies in our sample fall significantly below the size--mass relations at all redshifts, and all 12 galaxies meet the definition of compact massive galaxies from \citet{van15} (with 6/12 falling below their plot range of $r_e>0.2$~kpc). In terms of central stellar mass, the $r_{e,central}$ values are clearly extreme for the same 10/12 galaxies. As discussed in Section~\ref{sec:z2}, the physical resolution for rest-frame optical observations with HST is almost a factor of three better at $z\sim0.6$ compared to $z\sim2$. This observational effect and the small mass-to-light ratios for the central regions of our galaxies help explain their extreme compactness in terms of half-light radius.}
\label{fig:size_mass}
\end{center}
\end{figure*}

\subsection{Stellar Mass Surface Densities}\label{sec:sigma}

Here we consider the implications of the size measurements from Section~\ref{sec:photmorph} and central stellar mass estimates from Section~\ref{sec:stellarpop} for the stellar surface densities of these galaxies on $r\sim0.1$--1~kpc scales. The stellar surface density measured on these scales has been shown to be a clear way to differentiate star-forming galaxies from quiescent galaxies, suggesting that the quenching of star formation happens when a galaxy reaches a threshold in stellar surface density \citep[e.g.,][]{van15}. In addition, measurements of stellar surface density on $r=1$~kpc scales have been used to select compact star-forming galaxies that are likely in the process of becoming compact quiescent galaxies \citep[e.g.,][]{bar17,koc17}. 

We compute the following three measures of stellar surface density for each galaxy:
\begin{itemize}
    \item $\Sigma_{e,central}$: This is central stellar surface density within the half-light radius of the compact starburst component for each galaxy. Most galaxies in our sample have $r_{e,central}=0.1$--0.2~kpc, so this value is relevant for those size scales. Similar to Equation~\ref{eqn:esc}, this assumes that half of $\mstar{}_{,central}$ is within the half-light radius.
    \begin{equation}\label{eqn:sigma}
    \Sigma_{e,central} = \frac{\mstar{}_{,central}}{2 \pi r_{e,central}^{2}}
    \end{equation}
    \item $\Sigma_{e,total}$: This is calculated using $r_{e,total}$, which is the estimate of half-light radius for the entire galaxy from Section~\ref{sec:re_total}. This accounts for the extended emission in each galaxy (i.e., not just the compact starburst component), and the typical values are $r_{e,total}=0.2$--0.3~kpc (see Table~\ref{tab:re_tot}). %We use $\Sigma_{e,total}$ for subsequent comparisons in Section~\ref{sec:z2} and Figure~\ref{fig:sigma_mass}.
    \begin{equation}\label{eqn:sig_tot}
    \Sigma_{e,total} = \frac{\mstar{}_{,total}}{2 \pi r_{e,total}^{2}}
    \end{equation}
    \item $\Sigma_1$: This is the stellar surface density within a fixed aperture of $r=1$~kpc. Because the images on $r=1$~kpc (or $r=0.15$\arcsec) scales are still dominated by the light profile from the compact starburst component, it difficult to accurately measure the stellar mass of an additional component on these angular scales. We therefore determine a lower limit to the stellar mass within the central kpc by extrapolating the central light profiles out to $r=1$ kpc for each galaxy. Given that $r_{90}/r_{50}=5.55$ for an $n=4$ de Vaucouleurs profile and $r_{95}/r_{50}=8.64$ \citep{dev48,gra05}, this means that we are extrapolating from 50\% of the light on $r=0.1$--0.2~kpc scales to $90$--95\% of the light on $r=1$~kpc scales. In the equation below, $f$ represents the fraction of the central component's stellar mass that is contained within $r=1$~kpc (e.g., $f=95$\%). 
    \begin{equation}\label{eqn:sig1}
    \Sigma_{1} = \frac{\mstar{}_{,central}\times f}{\pi (1~\textnormal{kpc})^{2}}
    \end{equation}
\end{itemize}

We show the $\Sigma_{e,central}$ and $\Sigma_1$ values along with observed outflow velocities in Figure~\ref{fig:sigma_star}. We find a median value of $\Sigma_{e,central} =3\times10^{11}$~\sigunits, which matches the theoretical estimate of the Eddington limit found by \citet{hop10}. As discussed by \citet{hop10}, this is also comparable to the maximum stellar surface density for samples of $z\sim0$ elliptical galaxies \citep{lau07,kor09} and $z=2$--3 compact quiescent galaxies \citep{van08,bez09}, as well as lower-mass stellar systems such as M82 super star clusters \citep{mcc07} and the Milky Way nuclear disk \citep{lu09}. Accounting for uncertainties on both $r_e$ and central stellar mass, there are 10/12 galaxies in our sample (the only exceptions being J1558 and J1613, which we describe in Section~\ref{sec:sample}) that are consistent with $\Sigma_{e,central}=3\times10^{11}$~\sigunits. 

The values for $\Sigma_1$, estimated as lower limits as described above, are less extreme with a median value $\Sigma_{1}=7\times10^{9}$~\sigunits. Considering Equation~\ref{eqn:sigma} and Equation~\ref{eqn:sig1}, this is because the $r=1$~kpc area is larger than $r=r_e$ area by a factor of $\sim100$ in many cases, while the stellar mass of the central component within the $r=1$~kpc aperture is only larger by a factor of $\approx2$. The mean ratio is $\Sigma_{e,central}/\Sigma_1=50$ across the whole sample. We find no correlation between outflow velocity and either measure of central stellar surface density.

\subsection{Comparison to compact massive galaxies at $z\sim2$}\label{sec:z2}

The population of galaxies that we are studying may represent a brief, but common phase of galaxy evolution that many massive galaxies go through. Given the evidence that the number density of compact massive galaxies peaks at $z\approx2$ \citep[e.g.,][]{van14} and that many $\mstar\sim10^{11}$~M$_{\odot}$ quiescent galaxies have formed by this epoch \citep[e.g.,][]{mar09,ilb13,muz13,tom14}, it is worthwhile to compare to compact galaxies at $z=2$--3 that may be going through a similar phase. Regarding this comparison, an important point is that the population of compact starburst galaxies at $z\sim0.6$ from this paper is unlikely to be well characterized by existing surveys with HST. The follow-up sample of 131 galaxies described in Section~\ref{sec:sample} was selected over the 10,000~deg$^2$ of the SDSS-I survey, corresponding to an average density on the sky of $\sim0.01$ sources per square degree. For reference, the area of extragalactic fields like COSMOS \citep{sco07} and CANDELS \citep{gro11} are 1.8~deg$^2$ and 0.2~deg$^2$, respectively, such that there is a 2.3\% chance of one of these galaxies being in COSMOS and a 0.29\% chance of one of these galaxies being in CANDELS. 

To compare with the properties of compact star-forming galaxies and compact quiescent galaxies that are found in existing surveys, we show the relationship between stellar surface density (both $\Sigma_{e,total}$ and $\Sigma_1$) and stellar mass in Figure~\ref{fig:sigma_mass}. We include trends for quiescent and star-forming galaxies from \citet{bar17} at $0.5<z<1.0$, the redshift interval that overlaps with our sample, and $2.2<z<3.0$, the highest redshift interval available. There is cosmic evolution in these $\Sigma-\mstar$ relations, such that galaxies at higher redshift exhibit larger surface densities at a given stellar mass. In terms of $\Sigma_{e,total}$ all of the galaxies in our sample lie above the relation for quiescent galaxies at $0.5<z<1.0$. At $2.2<z<3.0$, there are quiescent galaxies from \citet{bar17} with $\Sigma_e$ values that approach $10^{11}$~\sigunits. The two galaxies from our sample with the smallest $\Sigma_{e,total}$ values (J1558, J1613, see Section~\ref{sec:sample} and Section~\ref{sec:sigma}) lie within the scatter of the $\Sigma_e-\mstar$ relation at $2.2<z<3.0$, and the remaining 10/12 all lie above the relation. In terms of $\Sigma_1$, there is good agreement between the lower-limit $\Sigma_1$ values that we calculated above (Section~\ref{sec:sigma}, Equation~\ref{eqn:sig1}) and the relation for $0.5<z<1.0$ quiescent galaxies. Two galaxies within our sample (J1341, J2116) extend slightly above the scatter at $0.5<z<1.0$, and are closer to the relation for $2.2<z<3.0$ quiescent galaxies. In total, 11/12 of our galaxies (all but J1558, the ongoing merger) would meet the $\Sigma_1$ compactness definition from \citet{bar17}. 

We show the relationship between half-light radius and stellar mass in Figure~\ref{fig:size_mass}, and we compare to early-type galaxies studied by \citet{van14}. In particular, we show the trends from \citet{van14} for galaxies at $0.5<z<1.0$ (the lowest redshift range in their sample, which also overlaps with our sample) and at $2.5<z<3.0$ (the highest redshift range in their sample, for which early-type galaxies are the most compact at a given stellar mass). When considering $r_{e,total}$ and total stellar mass, which is the most direct comparison to measurements made in other studies, 10/12 of the galaxies in our sample (all except J1558 and J1613) fall significantly below the size--mass relation for early-type galaxies at all redshifts. In particular, our galaxies have $r_{e,total}$ values that are below the size--mass relation at $2.5<z<3.0$ by a factor of 7 on average. In addition, all 12 galaxies meet the definition of compact massive galaxies from \citet{van15}, which is shown as the shaded gray region in Figure~\ref{fig:size_mass}. It is worth noting that three galaxies (J0905, J0944, J1341) have upper limits $r_{e,total}<0.2$~kpc such that they fall completely below the plot range used by \citet{van15}.

Based on our analysis of the mass-to-light ratios of the central and extended stellar components (Section~\ref{sec:stellarpop}), we know that the half-mass radii are significantly larger than the half-light radii for galaxies in our sample. Given how the light profiles are so dominated by the central component (Section~\ref{sec:photmorph}),  we have not attempted to measure detailed stellar mass profiles or to estimate half-mass radius (i.e., such an analysis would be challenging with the existing data and is beyond the scope of this paper). To isolate the properties of the compact starburst component in Figure~\ref{fig:size_mass}, we also show the values for $\mstar{}_{,central}$, for which $r_{e,central}$ is an accurate measure of the half-mass radius. The central stellar mass values for 6/12 galaxies in our sample are below $\mstar=2\times10^{10}~\msun$, which is as far as \citet{van14} extend their analysis. Among the remaining galaxies, 5/6 (all but J1613) still fall significantly below the size--mass relation at all redshifts. In addition, 2/12 galaxies also have a central stellar mass $\mstar>6\times10^{10}~\msun$, which exceeds the minimum stellar mass threshold adopted by \citet{van15} for their definition of compact massive galaxies. 

In this comparison to compact massive galaxies at z$=2$--3, one important point is that the angular resolution with HST at rest-frame $V$-band for galaxies at $z\sim0.6$ (e.g., $\textnormal{FWHM}\approx0.07\arcsec$ at F814W with WFC3/UVIS) is a factor of two better than it is for galaxies at $z\sim2$ (e.g., $\textnormal{FWHM}\approx0.15\arcsec$ at F160W for WFC3/IR). There is also a $25\%$ difference in terms of angular diameter distance (for our adopted $\Lambda$CDM cosmology), such that $r=0.2$~kpc is $0.030\arcsec$ at $z=0.6$ ($2.3\times$ smaller than the F814W FWHM) and $0.024\arcsec$ at $z=2$ ($6.3\times$ smaller than the F160W FWHM). Given the systematic issues with accurate characterization of the point-spread function, there is reason to be skeptical of any $r_e$ measurements on scales that are significantly smaller than the FWHM. If one adopted a floor of 20\% of the PSF size (i.e., stating that it is not possible to recover accurate $r_e$ measurements on scales smaller than this, and one should only quote upper limits), this corresponds to a limits of $r=90$~pc at $z=0.6$ and $r=250$~pc at $z=2$. In other words, it makes sense why \citet{van15} only extend down to $r=0.2$~kpc on their version of Figure~\ref{fig:size_mass}. This clear observational limit at $z\sim2$ thus precludes measurements of $r_e$ and $\Sigma_e$ that are as extreme as the values we find with better physical resolution at $z\sim0.6$. The way in which the galaxies in our sample are unique (e.g., in terms of $\Sigma_e$ in Figure~\ref{fig:sigma_star} and Figure~\ref{fig:sigma_mass}, and in terms of $r_e$ in Figure~\ref{fig:size_mass}) is related to both their physical compactness and our observational ability to recover information on $r<0.2$~kpc scales. While our estimates of $\Sigma_1$ using Equation~\ref{eqn:sig1} are lower limits, this comparison (e.g., in the bottom panels of Figure~\ref{fig:sigma_mass}) suggests that the galaxies in our sample, when measured on the same physical scales, have physical properties that are comparable to those of compact massive galaxies from the literature.

\subsection{Comparison with theoretical scenarios for the formation of compact starbursts}\label{sec:origin}

Previous work by \citet{sel14} used HST/F814W morphological information to show clear evidence that the galaxies in this sample formed via major mergers. That result is also apparent in the disturbed morphologies and tidal tails in Figure~\ref{fig:hst_images_1} and Figure~\ref{fig:hst_images_2}, particularly in the new F160W images. Furthermore, the compact morphology of the recent starburst component indicates there was significant dissipation of angular momentum for the cold gas component of these galaxies, most of which has subsequently been consumed or ejected \citep{gea13,gea14,gea18,rup19}. In that context, it is worth exploring how the properties of these galaxies compare to predictions from merger-driven evolutionary scenarios \citep[e.g.,][]{bar96,mih96,hop06,hop08,hay14} and models that funnel gas to the center of a galaxy to fuel nuclear star formation \citep[e.g.,][]{dek09,dek14,zol15,tac16}. 

Under the hypothesis that these are elliptical galaxies in formation \citep[e.g.,][]{too77,bar92,kor92}, the dominance of the light profiles by a compact starburst component suggests that we are witnessing the formation of ``power-law" ellipticals with central stellar cusps or ``extra light" at their centers \citep[e.g.,][]{fab97,lau07,kor09,hop09}. For example, \citet{hop09}, building on earlier work of \citet{mih94}, used merger simulations to analyze how the properties of the extra light in cusp ellipticals depend on stellar mass and gas fraction. In particular, by fitting the surface brightness profiles of local elliptical galaxies, \citet{hop09} estimated the fraction of the stellar mass formed in a central starburst $f_{sb}$. They found that $f_{sb}$ increases with gas fraction and decreases with stellar mass. For galaxies with $\mstar=10^{11}~\msun$, they found typical values of $f_{sb}\approx0.15$, and at $\mstar=10^{10}~\msun$ they found typical values of $f_{sb}\approx0.30$. In comparison, our result of $f_{sb}\approx0.25$ is somewhat above this trend, but it is consistent with the scatter found by \citet{hop09} for galaxies with $\mstar=10^{11}~\msun$.

Similar to the gas dissipation in major mergers that form compact starbursts, \citet{dek14} describe a model in which ``blue nuggets" are formed by violent disc instability that drives gas to the center of the galaxy. Subsequent studies using numerical simulations have explored both the onset of ``wet compaction" in this model and the subsequent quenching of star formation \citep[e.g.,][]{zol15,tac16}. In particular, \citet{dek14} argue that this instability happens when the timescale for inflow is shorter than the timescale for star formation, and they find that this happens when a threshold of $\approx28$\% of the mass in the disk is in a ``cold" component, which includes both gas and young stars. Given that we find $\approx25\%$ of the total stellar mass is in young stars for the galaxies in our sample, and given that the mass of molecular gas contributes an additional 3\%--15\% for these same galaxies \citep{gea13,gea14,gea18}, this clearly exceeds the threshold for violent disc instability. The large stellar surface densities that we measure are also all beyond the stellar surface density threshold of $\Sigma=10^9$~\sigunits, above which galaxies formed by violent disc instability are expected to dominate in this model. That said, based on the tidal features and evidence for recent merger activity in our sample, the compactness of the star formation that we observe is likely due to merger interactions.

\subsection{Comparison with models for driving fast outflows}\label{sec:fast}

As shown in Figure~\ref{fig:vesc} and discussed at the beginning of Section~\ref{sec:escape_surface} and in Section~\ref{sec:vesc}, our results are consistent with physical mechanisms that can drive outflows at speeds significantly faster than the central escape velocity. Here we discuss several models of star-formation feedback from the literature that can and cannot produce such fast outflows.

Based on a simple idealized model of a cloud that is accelerated outward by a source of momentum input $\dot{p}$ from launch radius $r_0$, \citet{hec11} describe how a physical configuration that includes a large number of massive stars within a small radius would produce a very large outward pressure and a fast outflow velocity. In this idealized model, the terminal velocity of the cloud is proportional to $\dot{p}^{1/2}$ (i.e., for momentum that comes from massive stars, this means that the outflow velocity depends on the square root of the star-formation rate) and proportional to $r_0^{-1/2}$ (i.e., a launch radius that is smaller by $\sim10\times$ would produce a cloud velocity that is larger by $\sim3\times$). For a fiducal value of $r_0\sim0.1$~kpc (and fiducial values for $\dot{p}$, cloud column density, and outflow opening angle) they show how this model can produce cloud velocities $v\sim2000$~\kms. This model applies for the limit in which the gravitational force is negligible in reducing a cloud's acceleration \citep[e.g.,][]{che85}, so it does not depend on the stellar mass, stellar surface density, or escape velocity; it only requires a large star-formation rate on $r\sim100$~pc scales.

Expanding on this idea of a momentum-driven population of clouds, \citet{hec15} introduce the parameter $R_{crit}$ to describe how $\dot{p}$ compares to the critical momentum flux required for the wind to overcome gravity acting on the clouds: $R_{crit}=\dot{p}/\dot{p}_{crit}$. They define strong outflows as having $R_{crit}>10$, and they find that galaxies in this regime typically have outflow velocities that exceed their $\sqrt{G M / r}$ circular velocities by a factor of $\sim3$. They also perform an analytic calculation for how the ratio of the maximum outflow velocity to the circular velocity would depend on $R_{crit}$ for this model, finding a relation that is roughly consistent with values estimated for a sample of 37 galaxies at $z<0.2$. \citet{hec16} extend this analysis to include 9 of the compact starburst galaxies from our HST sample \citep{dia12,sel14,gea14}, including $6/12$ of the galaxies from this paper. They argue that our $z\sim0.6$ galaxies lie on the same trends found for their sample of $z<0.2$ galaxies, extrapolated to larger values of outflow velocity, SFR, and $\sigmasfr$. It is worth noting that \citet{pet20} also find a correlation between outflow velocity and $\sigmasfr$ among 18 galaxies from our HST sample that have infrared- and radio-based estimates of SFR, including 9/12 galaxies from this paper.

In addition, there are analytic arguments \citep[e.g.,][]{tho15,bus16} and high-resolution simulations \citep[e.g.,][]{mur15,sch20} that provide potential explanations for the fast outflows that we observe. \citet{tho15} describe a model for shells and clouds driven by radiation pressure, and they find that the asymptotic velocity can significantly exceed $v_{esc}$, reaching speeds of 1000--2000~\kms. In this model, there is a phase as the shell propagates from sub-kpc scales to $\sim10$~kpc scales during which it is optically thick to UV photons that accelerate the gas to high velocities. \citet{bus16} rework the \citet{che85} model for supernova-driven winds to include non-uniform sources of mass and energy, extended gravitational potentials, and radiative losses. They find transonic wind solutions that can reproduce cool, fast outflows with speeds $\sim1000$~\kms. \citet{mur15} analyze galaxy-scale outflows in the Feedback in Realistic Environments \citep[FIRE,][]{hop14} simulations, which include radiation pressure, stellar winds, and ionizing feedback from young stars, along with energy and momentum input from supernovae that are resolved on the scale of giant molecular clouds. They find typical wind velocities that are 1--3$\times$ the circular velocty of the halo, but that the 95th percentile velocities approach 1000~\kms\ for the most massive halos in their sample ($\mhalo\sim10^{12}~\msun$). More recently, \citet{sch20} present results from the Cholla Galactic OutfLow Simulations \citep[CGOLS,][]{sch18a,sch18b} project, which include clustered supernova feedback and mixing of the hot and cool phases of the wind. Their simulations are set up to mimic the local starburst galaxy M82, which is an order of magnitude less massive than the galaxies in our sample \citep[e.g.,][]{gre12}, but they find cool outflows with speeds that approach 1000~\kms.

In contrast, models that predict that outflow velocity should be approximately equal to the escape velocity, circular velocity, or velocity dispersion of a galaxy are not consistent with our results. For example, \citet{opp06} implement a prescription for momentum-driven winds in their hydrodynamic simulations for which wind velocity is proportional to galaxy velocity dispersion, motivated by observational correlations \citep[e.g.,][]{mar05,rup05} and theoretical arguments \citep{mur05}. This momentum-driven wind model was used subsequently in simulations to reproduce the enrichment history of the intergalactic medium, the mass-metallicity relation, and galaxy stellar masses and star formation rates \citep[e.g.,][]{opp08,fin08,dav11}. A similar scaling in which wind velocity is equal to the local escape velocity has been implemented in semi-analytic models of galaxy formation \citep[e.g.,][]{dut10}. Furthermore, \citet{mur11} use analytic arguments to predict that wind velocity from radiation pressure should be similar to the escape velocity of the most massive star clusters in a galaxy. Such model predictions significantly underestimate the outflow velocities for galaxies in our sample. More recently, \citet{nel19} analyze the properties of galactic outflows from the IllustrisTNG project \citep[e.g.,][]{pil18,pil19}, including supernova-driven winds for which the outflow velocity is proportional to the local dark matter velocity dispersion. This model of stellar feedback is not consistent with the outflows we observe, and \citet{nel19} describe how black hole feedback associated with low-luminosity AGN activity is required to produce $>1000$~\kms\ wind velocities in their simulations.

In summary, both analytic and numerical models of pure starburst-driven winds are able to achieve the fast outflow velocities seen in our sample of galaxies, provided that those models include appropriate physical conditions associated with intense and compact star formation. However, hydrodynamic simulations that use sub-grid prescriptions based on correlations between outflow velocity and escape velocity or velocity dispersion will not reproduce these high velocities and may be missing some of the feedback energy injected by starbursts. There is clear motivation for further, more detailed comparisons between high-resolution simulations and the properties of the galaxies in our sample as part of future work. 

\section{Summary and Conclusions}\label{sec:summary}

We have focused in this paper on a sample of compact starburst galaxies that show clear evidence for recent merger activity and very fast outflows, and little evidence for ongoing AGN activity. We have confirmed that these $\mstar\sim10^{11}~\msun$ galaxies at $0.4<z<0.8$ have very compact sizes $r_{e,central}\sim100$~pc (Section~\ref{sec:uvis_re}), and we have measured for the first time the ultraviolet, optical, and infrared colors of the compact starburst component for each galaxy (Section~\ref{sec:galfitm}). Using this nuclear photometry as the basis for stellar population modeling (Section~\ref{sec:nucmass}), we find that the compact starburst component typically contributes 25\% of the total stellar mass (Section~\ref{sec:extended}). Based on the results for size and stellar mass, we find typical central escape velocities $v_{esc,central}\sim900$~\kms, which is a factor of two lower than the observed outflow velocities (Section~\ref{sec:vesc}, Figure~\ref{fig:vesc}). This requires physical mechanisms that can launch outflows at speeds significantly faster than the central escape velocity \citep[e.g.,][see Section~\ref{sec:fast}]{hec15,tho15}, and it indicates that these ionized outflows have sufficient momentum to escape the galaxy into the circumgalactic medium and potentially beyond \citep[e.g.,][]{gea14,rup19}. We also find central stellar densities $\Sigma_{e,central}\approx3\times10^{11}$~\sigunits\ that are comparable to theoretical estimates of the Eddington limit (Section~\ref{sec:sigma}, Figure~\ref{fig:sigma_star}). It is clear that the central stellar component in these galaxies formed recently (see Table~\ref{tab:sspmass} and Table~\ref{tab:burstmass} for age estimates) in an extremely dense episode of star formation that provided a large amount of momentum and energy input for driving the observed $|v|=1000$--3000~\kms\ outflows. 

We also estimate lower limits on $\Sigma_1$ stellar densities within the central kpc, and we compare $\Sigma_1$, $\Sigma_e$, and $r_e$ values to those of compact massive galaxies from the literature at $0.5<z<3.0$ (Section~\ref{sec:z2}). In terms of $\Sigma_e$ and $r_e$ at a given stellar mass, most galaxies in our sample have significantly higher surface densities and smaller half-light radii than even the most compact galaxies at $z=2$--3 (see Figure~\ref{fig:sigma_mass} and Figure~\ref{fig:size_mass}). This difference reflects both the dominance of a young central stellar component (i.e., the half-light radius is significantly smaller than the half-mass radius for galaxies in our sample), and the fact that rest-frame optical observations at $z\sim0.6$ have better physical resolution than rest-frame optical observations at $z\sim2$ (by a factor of $\approx2.7$, see Section~\ref{sec:z2}). When making comparisons on the same physical scales with $\Sigma_1$ estimates, we find values that are comparable to those of compact massive galaxies at $0.5<z<3.0$ (``blue nuggets" and ``red nuggets").

Taken as a whole, our results provide new information about the physical conditions at the centers of these galaxies and about the young stellar component that makes them so extreme. These findings are consistent with a scenario in which this central component formed by a process that funneled a large supply of gas to the center of the galaxy (e.g., a gas-rich major merger) and triggered compact star formation at high surface density near the Eddington limit, expelling most of the remaining gas supply in a fast outflow. Consistent with this picture, the high stellar surface densities and weak optical emission lines for these galaxies \citep[e.g., relative to their infrared luminosity,][]{dia12,sel14,pet20} suggest that they have begun the process of quenching and could be the progenitors of power-law or cusp ellipticals in the local universe that have prominent ``extra light" components (see Section~\ref{sec:origin}). There is clear motivation for future work to address open questions  about the recent history of star formation and AGN activity, the multi-phase structure and physical extent of the observed outflows, the physical processes responsible for ejecting the gas supply, and the space density and cosmic relevance of these galaxies, which may represent a short-lived pathway of ``compaction" and quenching for massive galaxies that is more common at $z=2$--3, but more observationally accessible in these rare examples at $z<1$.

\acknowledgments

We acknowledge support for HST-GO-13689 that was provided by NASA through grants from STScI to Bates College, Siena College, and the University of Wisconsin-Madison. We acknowledge support from the National Science Foundation (NSF) under a collaborative grant (AST-1813299, 1813365, 1814233, 1813702, and 1814159) and from the Heising-Simons Foundation grant 2019-1659. AMD acknowledges support from the Maine Space Grant Consortium through a grant to Bates College. CAT and JDD thank the UW-Madison H.I. Romnes Faculty Fellowship for support. We thank the referee for suggestions that have improved the paper. AMD had useful discussions about this project with Kwamae Delva, Sof\'ia Edgar, Kieran Edmonds, Will Jaekle, Natasha Jones, Du{\v{s}}an Kere{\v{s}}, Fahim Khan, Rebecca Minsley, Jos\'e Ruiz, Edgar Sarceno, Linn\'ea Selendy, Riley Theriault, Todd Thompson, and Emily Woods, and received helpful support from Theresa Bishop, Shonna Humphrey, Alison Keegan, Kerry O'Brien, Joseph Tomaras, and Heather Ward. The National Radio Astronomy Observatory is a facility of the National Science Foundation operated under cooperative agreement by Associated Universities, Inc.  


\vspace{5mm}
\facilities{HST(WFC3)}


\software{\astropy\ \citep{ast13,ast18}, \dynesty\ \citep{spe20}, \galfit\ \citep{pen02,pen10}, \galfitm\ \citep{hau13,vik13}, \fsps\ \citep{con09,con10}, \prospector\ \citep{lej17,joh19} }




\bibliography{ads}

\begin{deluxetable*}{ccrrcc}
\tablecaption{Results from central SSP modeling \label{tab:sspmass}} 
\tablehead{
\colhead{name} & \colhead{$\log(\mstar [\msun])$} & \colhead{16th percentile} & \colhead{84th percentile} & \colhead{$t_{age}$ [Myr]} & \colhead{adopt}
}
\decimalcolnumbers
\startdata
J0826 & 10.27 & 10.23 ($-0.04$) & 10.32 ($+0.05$) & 29--35 & Y \\
J0901 & 10.11 & 10.05 ($-0.06$) & 10.16 ($+0.05$) & 28--36 & Y \\ 
J0905 & 10.39\tablenotemark{a} & 10.35 ($-0.04$) & 10.42 (+0.03) & 29--34 & N \\ 
J0944 & 10.20 & 10.04 ($-0.16$) & 10.31 ($+0.11$) & 19--32 & Y \\ 
J1107 & 10.04* & 10.01 ($-0.03$) & 10.52 ($+0.48$) & 3.3--35 & N \\ 
J1219 & 10.45 & 10.38 ($-0.07$) & 10.55 ($+0.10$) & 33--67 & Y \\ 
J1341 & 10.59* & 10.46 ($-0.13$) & 10.66 ($+0.07$) & 3.1--3.5 & N \\ 
J1506 & 10.75* & 10.68 ($-0.07$) & 10.80 ($+0.05$) & 3.1--3.4 & N \\ 
J1558 & 9.85 & 9.74 ($-0.11$) & 9.96 ($+0.11$) & 19--29 & Y \\ 
J1613 & 10.65 & 10.52 ($-0.13$) & 10.72 ($+0.07$) & 26--39 & Y \\ 
J2116 & 9.84* & 9.76 ($-0.08$) & 10.36 ($+0.52$) & 4.3--33 & N \\ 
J2140 & 10.56 & 10.49 ($-0.07$) & 10.66 (+0.10) & 33--63 & Y \\ 
\enddata
\tablecomments{Column 1: Short SDSS name (see Table~\ref{tab:sample}). Columns 2: Median (50th percentile) of the posterior probability distribution for stellar mass under the assumption of a single-age, simple stellar population. Column 3: Lower limit of the 68\% confidence interval for central stellar mass based on the 16th percentile of the posterior probability distribution. The quantity in parentheses shows the difference from the median value in dex. Column 3: Upper limit of the 68\% confidence interval for central stellar mass based on the 84th percentile of the posterior probability distribution. The quantity in parentheses shows the difference from the median value in dex. Column 4: The 68\% confidence interval on stellar age (16th percentile to 84th percentile of the posterior) under this assumption of a simple stellar population. Column 5: A flag of ``yes" or ``no" regarding whether we adopt the central stellar mass in this table for our subsequent analysis. For galaxies flagged as N, we use the central stellar mass reported in Table~\ref{tab:burstmass}.}
\tablenotetext{a}{As discussed in Section~\ref{sec:nucmass} and shown in Figure~\ref{fig:bursty_sed}, the nuclear SED for this galaxy is not fit robustly by single-age SSP models. This galaxy is also flagged as N in Column 6.}
\end{deluxetable*}

\begin{deluxetable*}{ccrrccc}
\tablecaption{Results from central delayed-$\tau$ modeling with late-time bursts \label{tab:burstmass}} 
\tablehead{
\colhead{name} & \colhead{$\log(\mstar [\msun])$} & \colhead{16th percentile} & \colhead{84th percentile} & \colhead{$t_{age}$ [Myr]}
& \colhead{$f_{burst}$} & \colhead{$f_{age,burst}$} }
\decimalcolnumbers
\startdata
J0905 & 10.41 & 10.17 ($-0.24$) & 10.70 (+0.29) & 6.6--25 & 0.24--0.78 & 0.79--0.97 \\ 
J1107 & 10.11 & 9.79 ($-0.32$) & 10.44 ($+0.33$) & 4.4--20 & 0.26--0.81 & 0.78--0.96 \\ 
J1341 & 10.51 & 10.26 ($-0.25$) & 10.66 ($+0.15$) & 3.8--10 & 0.22--0.80 & 0.65--0.92 \\
J1506 & 10.18 & 9.98 ($-0.20$) & 10.39 ($+0.21$) & 11--26 & 0.28--0.77 & 0.84--0.97 \\ 
J2116 & 10.68 & 10.42 ($-0.26$) & 10.83 ($+0.15$) & 3.8--8.8 & 0.21--0.82 & 0.62--0.90 \\ 
\enddata
\tablecomments{Column 1: Short SDSS name (see Table~\ref{tab:sample}). Columns 2: Median (50th percentile) of the posterior probability distribution for central stellar mass for the model assumption of a delayed-$\tau$ star-formation history with a late-time burst of star formation (see Section~\ref{sec:nucmass}). Column 3: Lower limit of the 68\% confidence interval for central stellar mass based on the 16th percentile of the posterior probability distribution. The quantity in parentheses shows the difference from the median value in dex. Column 4: Upper limit of the 68\% confidence interval for central stellar mass based on the 84th percentile of the posterior. The quantity in parentheses shows the difference from the median value in dex. Column 5: The 68\% confidence interval on stellar age (16th percentile to 84th percentile of the posterior), which for this model refers to the age of the oldest stars that contribute to the central stellar population. Column 6: The 68\% confidence interval on the fraction of the stellar mass that comes from the late-time burst. Column 7: The 68\% confidence interval for the time of the late-time burst of star formation. This is expressed as a fraction of $t_{age}$ in the sense that 0.5 refers to halfway through the star-formation history (i.e., a burst that occurred $t_{age}/2$ years ago) and 1 refers to the end of the star-formation history (i.e., a burst that is occurring at the most recent epoch).}
\end{deluxetable*}

\begin{deluxetable*}{ccrrcccc}
\tablecaption{Total emission: results from delayed-$\tau$ modeling with late-time bursts \label{tab:tot_params}} 
\tablehead{
\colhead{name} & \colhead{$\log(\mstar [\msun])$} & \colhead{16th percentile} & \colhead{84th percentile} & \colhead{$t_{age}$ [Gyr]} & \colhead{$\tau$ [Gyr]}
& \colhead{$f_{burst}$} & \colhead{$f_{age,burst}$} }
\decimalcolnumbers
\startdata
J0826 & 10.90 & 10.87 ($-0.03$) & 10.96 ($+0.06$) & 3.0--5.8 & 0.14--0.74 & 0.39--0.48 & 0.9915--0.9956 \\
J0901 & 10.81 & 10.78 ($-0.03$) & 10.86 ($+0.05$) & 3.4--5.8 & 0.14--0.50 & 0.40--0.49 & 0.9925--0.9956 \\
J0905 & 10.98 & 10.95 ($-0.03$) & 11.03 ($+0.05$) & 3.7--6.1 & 0.14--0.62 & 0.40--0.48 & 0.9911--0.9946 \\
J0944 & 10.80 & 10.75 ($-0.05$) & 10.86 ($+0.06$) & 1.3--2.7 & 0.17--7.9 & 0.38--0.48 & 0.9804--0.9902 \\
J1107 & 10.89 & 10.85 ($-0.04$) & 10.93 ($+0.04$) & 2.6--5.0 & 0.14--0.54 & 0.42--0.49 & 0.9869--0.9933 \\
J1219 & 11.11 & 11.06 ($-0.05$) & 11.17 ($+0.06$) & 1.2--2.4 & 0.27--9.0 & 0.33--0.47 & 0.9720--0.9848 \\
J1341 & 10.86 & 10.84 ($-0.02$) & 10.90 ($+0.04$) & 4.5--6.5 & 0.13--0.46 & 0.43--0.49 & 0.9941--0.9960 \\
J1506 & 10.84 & 10.80 ($-0.04$) & 10.89 ($+0.05$) & 1.2--2.0 & 0.32--11 & 0.35--0.48 & 0.9805--0.9884 \\
J1558 & 10.77 & 10.72 ($-0.05$) & 10.83 ($+0.06$) & 1.1--1.9 & 0.47--13 & 0.24--0.41 & 0.9794--0.9871 \\
J1613 & 11.13 & 11.09 ($-0.04$) & 11.18 ($+0.05$) & 1.1--1.6 & 1.0--14 & 0.14--0.24 & 0.9800--0.9858 \\
J2116 & 11.11 & 11.04 ($-0.07$) & 11.20 ($+0.09$) & 1.2--3.2 & 0.28--12 & 0.33--0.47 & 0.9671--0.9870 \\
J2140 & 11.16 & 11.10 ($-0.06$) & 11.21 ($+0.05$) & 1.3--2.4 & 0.16--5.5 & 0.41--0.49 & 0.9738--0.9855 \\
\enddata
\tablecomments{Column 1: Short SDSS name (see Table~\ref{tab:sample}). Columns 2: Median (50th percentile) of the posterior probability distribution for total stellar mass for the model assumption of a delayed-$\tau$ star-formation history with a late-time burst of star formation (see Section~\ref{sec:nucmass}). Column 3: Lower limit of the 68\% confidence interval for total stellar mass based on the 16th percentile of the posterior probability distribution. The quantity in parentheses shows the difference from the median value in dex. Column 4: Upper limit of the 68\% confidence interval for total stellar mass based on the 84th percentile of the posterior. The quantity in parentheses shows the difference from the median value in dex. Column 5: The 68\% confidence interval on stellar age (16th percentile to 84th percentile of the posterior), which for this model refers to the age of the oldest stars that contribute to the total stellar population. Column 6: The 68\% confidence interval on the $\tau$ e-folding parameter for the star-formation history in Gyr. Column 7: The 68\% confidence interval on the fraction of the total stellar mass that comes from the late-time burst in this model. Column 8: The 68\% confidence interval for the time of the late-time burst of star formation. This is expressed as a fraction of $t_{age}$ in the sense that 0.5 refers to halfway through the star-formation history (i.e., a burst that occurred $t_{age}/2$ years ago) and 1 refers to the end of the star-formation history (i.e., a burst that is occurring at the most recent epoch).}
\end{deluxetable*}

\begin{deluxetable*}{ccrrccc}
\tablecaption{Extended emission: results from delayed-$\tau$ modeling with late-time bursts \label{tab:resid}} 
\tablehead{
\colhead{name} & \colhead{$\log(\mstar [\msun])$} & \colhead{16th percentile} & \colhead{84th percentile} & \colhead{$t_{age}$ [Gyr]}
& \colhead{$f_{burst}$} & \colhead{$f_{age,burst}$} }
\decimalcolnumbers
\startdata
J0826 & 10.79 & 10.66 ($-0.13$) & 10.94 ($+0.15$) & 1.3--4.0 & 0.10--0.42 & 0.61--0.91 \\  
J0901 & 10.69 & 10.57 ($-0.12$) & 10.81 ($+0.12$) & 1.2--3.3 & 0.10--0.41 & 0.61--0.92 \\
J0944 & 10.56 & 10.44 ($-0.12$) & 10.69 ($+0.13$) & 1.3--3.7 & 0.14--0.44 & 0.63--0.90 \\
J1107 & 10.74 & 10.66 ($-0.08$) & 10.84 ($+0.10$) & 1.3--3.9 & 0.10--0.41 & 0.61--0.90 \\
J1219 & 11.03 & 10.94 ($-0.09$) & 11.12 ($+0.09$) & 1.4--4.3 & 0.12--0.42 & 0.60--0.89 \\
J1558 & 10.77 & 10.70 ($-0.07$) & 10.84 ($+0.07$) & 1.2--2.8 & 0.08--0.37 & 0.61--0.92 \\
J2116 & 11.11 & 11.00 ($-0.11$) & 11.25 ($+0.14$) & 1.4--4.1 & 0.11--0.41 & 0.61--0.91 \\ 
\enddata
\tablecomments{Column 1: Short SDSS name (see Table~\ref{tab:sample}). Columns 2: Median (50th percentile) of the posterior probability distribution for extended stellar mass for the model assumption of a delayed-$\tau$ star-formation history with a late-time burst of star formation (see Section~\ref{sec:nucmass}). Column 3: Lower limit of the 68\% confidence interval for extended stellar mass based on the 16th percentile of the posterior probability distribution. The quantity in parentheses shows the difference from the median value in dex. Column 4: Upper limit of the 68\% confidence interval for extended stellar mass based on the 84th percentile of the posterior. The quantity in parentheses shows the difference from the median value in dex. Column 5: The 68\% confidence interval on stellar age (16th percentile to 84th percentile of the posterior), which for this model refers to the age of the oldest stars that contribute to the extended stellar population. Column 6: The 68\% confidence interval on the fraction of the extended stellar mass that comes from the late-time burst in this model. Column 7: The 68\% confidence interval for the time of the late-time burst of star formation. This is expressed as a fraction of $t_{age}$ in the sense that 0.5 refers to halfway through the star-formation history (i.e., a burst that occurred $t_{age}/2$ years ago) and 1 refers to the end of the star-formation history (i.e., a burst that is occurring at the most recent epoch).}
\end{deluxetable*}

\end{document}